\documentclass[%
readprint,
twocolumn,
groupedaddress,
showpacs,preprintnumbers,
nofootinbib,
aps,
prd,
showkeys
]{revtex4-1}

\usepackage{subfigure}

\usepackage{amsmath}
\usepackage{amssymb}
\usepackage{mathtools}
\usepackage{hyperref}
\usepackage{placeins}
\usepackage{graphicx}
\usepackage{dcolumn}

\usepackage{color}

\usepackage{ulem}

\usepackage{comment}

\usepackage{float}
\usepackage{tabularx}
\usepackage{multirow}
\usepackage[html]{xcolor}
\usepackage{booktabs}
\newcolumntype{d}[1]{D{.}{.}{#1}}

\usepackage{upgreek}

\newcommand{\Rc}{R_\text{c}} 
\newcommand{\Rstar }{R_*} 
\newcommand{\Tc}{T_\text{c}} 
\newcommand{\vw}{v_\text{w}} 
\newcommand{\vwass}{v_\mathrm{w, late}} 
\newcommand{\lw}{l_\mathrm{w}} 

\newcommand{\ts}{\mathcal{T}} 
\newcommand{\xiw}{\xi_\text{w}} 
\newcommand{\TN}{T_\text{n}} 
\newcommand{\Rb}{R_\mathrm{b}} 
\newcommand{\Rd}{R_\mathrm{d}} 
\newcommand{\RdI}{R_{\mathrm{d}0}} 

\newcommand{\vws}{\left|\vw\right|} 

\newcommand{\Kd}{K_\mathrm{d}} 
\newcommand{\Kb}{K_\mathrm{b}} 
\newcommand{\Rbref}{R_\mathrm{ref}} 
\newcommand{\Kbref}{\Kb(\Rbref)} 

\newcommand{\Ubarf}{\overline{U}_\mathrm{f}} 
\newcommand{\Ubarfb}{\overline{U}_\mathrm{f,b}} 

\newcommand{\cs}{c_\mathrm{s}} 
\newcommand{\cJ}{c_\mathrm{J}} 

\newcommand{\Str}{\alpha} 

\newcommand{\phib}{\phi_\mathrm{b}} 

\newcommand{\potZeroT}{V_\text{0}} 
\newcommand{\potT}{V} 

\graphicspath{
  {./}
  {./figures/}
  {./figures/3d_sim_slices/}
  {./figures/3d_wall_speed_plots/}
  {./figures/similarity_curves/}
  {./figures/droplet_sketch/}
  {./figures/lineouts/}
  {./figures/rbtdot_K_wrt_time/}
  {./figures/K_scatter/}
  {./figures/similarity_sim_comp/}
  {./figures/3d_wall_speed_plots/}
  {./figures/vw_late/}
  {./figures/Rd0_comparison/}
  {./figures/dabya_comparison/}
}

\begin{document}

\newcommand{\HIPetc}{\affiliation{
Department of Physics and Helsinki Institute of Physics,
PL 64,
FI-00014 University of Helsinki,
Finland
}}

\title{Droplet collapse during strongly supercooled transitions}

\author{Daniel Cutting}
\email{daniel.cutting@helsinki.fi}
\author{Essi Vilhonen}
\email{essi.vilhonen@helsinki.fi}
\author{David J. Weir}
\email{david.weir@helsinki.fi}
\HIPetc

\date{\today}

\preprint{HIP-2022-5/TH}

\begin{abstract}
  We simulate the decay of isolated, spherically symmetric droplets in
  a cosmological phase transition. It has long been posited that such
  heated droplets of the metastable state could form, and they have
  recently been observed in 3D multi-bubble simulations. In those
  simulations, the droplets were associated with a reduction in the
  wall velocity and a decrease in the kinetic energy of the fluid,
  with a consequent suppression in the gravitational wave power
  spectrum. In the present work, we track the wall speed and kinetic
  energy production in isolated droplets and compare them to those
  found in multi-bubble collisions. The late-time wall velocities that
  we observe match those of the 3D simulations, though we find that
  the spherical simulations are a poor predictor of the kinetic energy
  production.  This implies that spherically symmetric simulations
  could be used to refine baryogenesis predictions due to the
  formation of droplets, but not to estimate any accompanying
  suppression of the gravitational wave signal.
\end{abstract}
\maketitle

\section{Introduction}

Many well-motivated extensions to the Standard Model have one or more
cosmological phase transitions as a feature (see
e.g.~\cite{Mazumdar:2018dfl,Hindmarsh:2020hop} and references
therein). If such a phase transition is of first order, it can have
interesting phenomenological consequences, such as the production of a
baryon asymmetry~\cite{Morrissey:2012db}, the generation of
gravitational waves~\cite{Weir:2017wfa,Caprini:2019egz}, or the
seeding of intergalactic magnetic
fields~\cite{Durrer:2013pga,Vachaspati:2020blt}.

With upcoming gravitational wave detectors like LISA offering enhanced
observational prospects for cosmological gravitational wave
backgrounds, there have been increased efforts to understand
first-order phase transitions in precise detail. Recent years have
seen advances in the determination of the asymptotic wall speed for
expanding
bubbles~\cite{Bodeker:2017cim,Hoche:2020ysm,Azatov:2020ufh,Laurent:2020gpg,Bea:2021zsu,Gouttenoire:2021kjv,Dorsch:2021nje,DeCurtis:2022hlx,Bea:2022mfb},
more precise calculations of the thermodynamic phase transition
parameters and nucleation
rates~\cite{Gould:2019qek,Kainulainen:2019kyp,Croon:2020cgk,Gould:2021ccf,Croon:2021vtc,Gould:2021oba,Hirvonen:2021zej,Lofgren:2021ogg,Ekstedt:2021kyx,Ekstedt:2022tqk},
and refined baryogenesis
computations~\cite{Cline:2020jre,Azatov:2021irb,Baldes:2021vyz,Dorsch:2021ubz,Lewicki:2021pgr}. Holographic
techniques have been used to compute phase transition parameters and
gravitational wave
signals~\cite{Ares:2020lbt,Bigazzi:2020phm,Bea:2021zsu,Bea:2021zol,Ares:2021ntv,Bea:2022mfb}. Preliminary
studies have explored the ability of LISA to reconstruct phase
transition parameters~\cite{Gowling:2021gcy,Giese:2021dnw}.  New
simulation techniques~\cite{Jinno:2020eqg}, valid for weak and
intermediate thermal transitions, have enabled the exploration of the
effects of density perturbations on the gravitational wave
spectrum~\cite{Jinno:2021ury}. The first simulations of gravitational
wave production from (magneto)hydrodynamic turbulence (including
acoustic turbulence) have been
conducted~\cite{RoperPol:2019wvy,Kahniashvili:2020jgm,RoperPol:2021xnd,Brandenburg:2021bvg,Dahl:2021wyk,RoperPol:2022iel},
and the gravitational wave signal from both strong thermal phase
transitions~\cite{Cutting:2019zws,Jinno:2019jhi} and vacuum-like
transitions is being
explored~\cite{Cutting:2018tjt,Cutting:2020nla,Lewicki:2020jiv,Lewicki:2020azd,Lewicki:2021xku,Gould:2021dpm}.
 
In Ref.~\cite{Cutting:2019zws}, hot droplets of the metastable state
were observed to form for strong transitions. This only occurred when
the reaction front was a deflagration, in which the fluid is
accelerated and heated ahead of the phase boundary. These hot droplets
consist of relatively small regions of the metastable state (false
vacuum) that are heated to well above the nucleation temperature, with
the resulting pressure opposing the progress of the phase boundary.

The formation of droplets in a cosmological transition is not a new
idea, and has been studied previously, primarily in the context of a
QCD-like phase transition. Early works on cosmological phase
transitions posited that, in the small-supercooling limit, transitions
that proceeded via deflagrations would reheat a substantial fraction
of the universe up to the critical
temperature~\cite{Witten:1984rs,Alcock:1985vc,Kajantie:1986hq,Alcock:1988br,Kurki-Suonio:1988wuo,Olesen:1991zt,Ignatius:1993qn,Heckler:1994uu}.
If such reheating were to take place, the growth of the bubbles would
slow, and the final stages of the transition would involve the
contraction of hot droplets of the metastable state. It was also
argued that baryons could become trapped inside these shrinking
droplets, leaving behind a baryon inhomogeneity on evaporation, or
even resulting in persistent so-called `nuggets' of the metastable
state.  If axions are present during the transition, this can modify
the stability of quark nuggets, leading to axion quark
nuggets~\cite{Hindmarsh:1991ay,Zhitnitsky:2002qa,Ge:2019voa,Zhitnitsky:2021iwg}. Furthermore,
it has been suggested that this mechanism could have taken place in a
dark sector~\cite{Bai:2018dxf}, offering another scenario in which
nuggets could comprise dark matter.

Other early works investigated the decay of spherical droplets using
spherically symmetric simulations and found that the fluid evolution
exhibited self-similar
behaviour~\cite{Rezzolla:1995kv,Rezzolla:1995br,Kurki-Suonio:1995yaf}. Spherical
simulations of expanding bubbles in the small-supercooling limit with
reflective boundary conditions were studied in
Ref.~\cite{Kurki-Suonio:1996gkq}. The reflective boundary was intended
to model the effect of interactions with `neighbouring bubbles'. When
the compression wave from deflagrations collided with the boundary, it
was found that the metastable region was reheated to the critical
temperature at which the two phases become degenerate.  As it cooled,
the remaining metastable region subsequently collapsed with a
substantially slower wall velocity than before.

In light of the above, the novelty of Ref.~\cite{Cutting:2019zws} was
that droplets were observed in three-dimensional (3D) simulations with
multiple bubbles, and away from the small-supercooling
limit. Furthermore, it was seen that they were associated with a
reduction in the expected kinetic energy fraction, which resulted in a
suppression of the gravitational wave signal. The exact physical
mechanism of this relationship was unclear.

In this paper, we revisit the formation and decay of droplets in the
case of spherical symmetry. We focus on transitions with an
expanding-bubble asymptotic wall speed of $\xiw=0.24$, as the droplets
formed in these cases were the longest lasting in
Ref.~\cite{Cutting:2019zws}. We perform a series of simulations of
collapsing droplets with a range of transition strengths, moving away
from the small-supercooling limit considered in earlier works. We
track the wall velocity and measure the kinetic energy production of
the droplets, comparing it to the multi-bubble results found in
Ref.~\cite{Cutting:2019zws}.

The layout of the paper is as follows. In
Section~\ref{sec:field-fluid}, we review the coupled field--fluid
model used to model the phase transition. In
Section~\ref{sec:dynamics}, we discuss the dynamics of expanding
bubbles and shrinking droplets in first-order phase transitions,
including similarity solutions and kinetic energy production.  We
describe our simulation code and the initial conditions in
Section~\ref{sec:compmethods}. We analyse the wall velocity and
kinetic energy production of our simulations in
Section~\ref{sec:Results}, and discuss similarity solutions in the
context of our results. We conclude in
Section~\ref{sec:Conclusions}. In the appendices, we provide a short
description of the wall speed estimators used in this paper in
Appendix~\ref{app:wallspeed}, then investigate the effect of varying
the fractional change in the number of degrees of freedom and initial
droplet radius in Appendices~\ref{app:fractionalchange}
and~\ref{app:convergence}, respectively.
 
\section{Coupled field--fluid model}
\label{sec:field-fluid}

Hydrodynamical simulations of phase transitions often employ the
coupled field--fluid model, in which a real scalar field is coupled to
a perfect fluid via a dissipative friction
term~\cite{Enqvist:1991xw,Ignatius:1993qn}.

The energy--momentum tensor of the coupled field--fluid model is given
by
\begin{equation}
  T^{\mu\nu} = (\epsilon + p)U^{\mu}U^{\nu} + p g^{\mu\nu} + \partial^{\mu} \phi \partial^{\nu} \phi - \frac{1}{2} g^{\mu\nu} \partial_\rho \phi\partial^\rho \phi\text,
\end{equation}
where $\epsilon$ and $p$ are the internal energy density and pressure
of the fluid, $\phi$ is the order parameter of the transition, and
$U=\gamma(1,\mathbf{v})$ with $\mathbf{v}$ the fluid 3-velocity and
$\gamma$ the associated Lorentz factor. The enthalpy of the system is
$w=\epsilon + p$.

The energy--momentum tensor can be split non-uniquely into a field and
fluid piece, such that $T^{\mu\nu}=T^{\mu\nu}_{\phi} +
T^{\mu\nu}_\mathrm{f}$. We make the choice that
\begin{align} 
  T^{\mu\nu}_\mathrm{f} &= (\epsilon + p)U^{\mu}U^{\nu} + g^{\mu\nu} p + \potT g^{\mu\nu}\text, \label{eq:Tf} \\ 
  T^{\mu\nu}_\mathrm{\phi} &= \partial^{\mu} \phi \partial^\nu \phi - g^{\mu\nu} (\partial \phi)^2 - \potT g^{\mu\nu}\text,
                             \label{eq:Tphi}
\end{align}
where $\potT$ is the effective thermal potential. We then assume that
the interaction between the field and the fluid can be modelled
via a phenomenological friction term
\begin{equation} \label{eq:coupling}
\partial_\mu T^{\mu\nu}_\phi = - \partial_\mu T^{\mu\nu}_\mathrm{f} = \eta U^\mu \partial_\mu \phi \partial^\nu \phi \text, 
\end{equation}
where $\eta$ is some constant friction parameter that is then set by
the particle physics theory in question. In principle, $\eta$ can be
derived from the microphysics of the phase transition and may depend
upon the order parameter and thermodynamic quantities in the vicinity
of the bubble wall (or, more generally, the phase
boundary)~\cite{John:2000zq,Konstandin:2014zta,Dorsch:2018pat,Laurent:2020gpg,Dorsch:2021nje}.
However, in this study we consider a simplified model in which we
treat $\eta$ as a constant free parameter.

An equation of state is needed to complete the field--fluid
system. Following Ref.~\cite{Cutting:2019zws}, we use a bag-like
equation of state:
\begin{align}
  \epsilon(T,\phi) &= 3a(\phi) T^4 + \potZeroT(\phi)\text, \\
p(T,\phi) &= a(\phi) T^4 - \potZeroT(\phi)\text,
\end{align}
where the zero-temperature effective potential is given by
\begin{equation}
  \potZeroT(\phi) = \frac{1}{2} M^2 \phi^2 + \frac{1}{3} \mu
\phi^3 + \frac{1}{4} \lambda \phi^4 - V_\mathrm{c}\text.
\end{equation}
Here $V_\mathrm{c}$ is a constant chosen such that the
zero-temperature potential is normalised to $\potZeroT(\phib)=0$,
where $\phib$ is the value of the scalar field in the broken phase. We
denote the potential energy difference at zero temperature with
$\Delta \potZeroT=\potZeroT(0) - \potZeroT(\phib)$.  This choice of
equation of state and effective potential is made for consistency with
Ref.~\cite{Cutting:2019zws}, where these droplets were first observed
in a three-dimensional simulation.

The temperature-dependent potential in this model is then given by
\begin{equation}
  \potT(\phi,T)=\potZeroT(\phi) - T^4\left(a(\phi)-a_0\right),
\end{equation}
where $a_0=(\pi^2/90)g_*$ and $g_*$ is number of the effective degrees
of freedom in the symmetric phase.

The function $a(\phi)$ models the change in the effective degrees of
freedom during the transition. We choose the form
\begin{equation}
a(\phi) = a_0 - \frac{\Delta \potZeroT}{\Tc^4} \left[ 3 \left(\frac{\phi}{\phib}\right)^2 -2\left(\frac{\phi}{\phib}\right)^3\right]\text.
\end{equation}
This form is convenient as it ensures that the minima of $\potZeroT$
at $\phi=0$ and $\phi=\phib$ remain the minima of $\potT$ for all
temperatures $T$. The change in effective degrees of freedom is given by\footnote{Note in some references $a_0$, the degrees of freedom in the symmetric phase, is denoted $a_+$. Similarly $a(\phib)$ is sometimes written as $a_-$.}
  \begin{equation}
    \label{eq:delta_a}
    \Delta a \equiv a_0 - a(\phib) = \frac{\Delta \potZeroT}{\Tc^4}.
  \end{equation}
Furthermore,
the two minima of the potential become degenerate at
$T=\Tc$, which is referred to as the critical temperature. Note that
with our equation of state, the speed of sound
\begin{equation}
\cs = \sqrt{\frac{\mathrm{d}p}{\mathrm{d}\epsilon}}
\end{equation}
is simply that of a relativistic fluid, $\cs^2=1/3$, in both
phases. Although convenient, these choices represent a
  significant simplification. This simplification was made in previous
  work to ensure numerical stability when exploring systems with large
  supercooling. We defer exploration of more realistic effective
  potentials and equations of state to future work.

To describe the phase transition, we define a phase transition
strength which measures the relative energy released during the
transition with respect to the radiation energy already in the plasma,
\begin{equation}
  \Str = \frac{\theta(0,\TN) - \theta(\phib,\TN)}{\epsilon_\mathrm{r}(\TN)}\text.
\end{equation}
Here $\TN$ is the nucleation temperature and $\epsilon_\mathrm{r} =
3w/4$ is the radiation energy density in the symmetric phase, in this
case $\epsilon_\mathrm{r} = 3a_0 T_n^4$.  The trace anomaly $\theta$
is given by
\begin{equation}
\theta(\phi,T) = \frac{1}{4}(\epsilon(\phi,T) - 3p(\phi, T))\text.
\end{equation}

Assuming spherical symmetry, we can derive the equation of motion for
the scalar field by considering $\partial_\mu T^{\mu\nu}_\phi$:
\begin{equation} \label{eq:phi-EOM}
- \ddot{\phi} + \frac{1}{r^2} \partial_r(r^2 \partial_r \phi) - \frac{\partial V}{\partial \phi} = \eta \gamma(\dot{\phi} + v \partial_r \phi)\text.
\end{equation}
Here $v$ is positive when the fluid velocity is pointed radially
outward.  Equations of motion for our other dynamical variables, the
fluid energy density $E=\gamma \epsilon$ and the fluid momentum
density $Z=\gamma^2 w v$, can be derived from $\partial_\mu
T^{\mu\nu}_\mathrm{f}$:
\begin{align}
  \dot{E} &+ \frac{1}{r^2} \partial_r (r^2 E v) + p\left[ \dot{\gamma} + \frac{1}{r^2} \partial_r (r^2 \gamma v)\right] \nonumber \\ 
          &- \frac{\partial{\potT}}{\partial \phi} \gamma(\dot{\phi} + v \partial_r \phi) = \eta \gamma^2 (\dot{\phi} + v \partial_r \phi)^2\text, \label{eq:E-EOM} \\
  \dot{Z} &+ \frac{1}{r^2} \partial_r (r^2 Z v) + \partial_r p + \frac{\partial \potT}{\partial \phi} \partial_r \phi \nonumber \\
          &= -\eta \gamma (\dot\phi + v \partial_r \phi)\partial_r \phi\text. \label{eq:Z-EOM}
\end{align}
In both cases we have taken the friction term of Eq.~\ref{eq:coupling}
into account.  These equations can then be discretised and solved
numerically, see Section~\ref{sec:compmethods}.

\section{Hydrodynamics of expanding bubbles and shrinking droplets}
\label{sec:dynamics}

In a thermal first-order phase transition, bubbles of the true vacuum
nucleate in the presence of a cosmic plasma. As the bubbles expand,
the friction between the bubble wall and plasma causes a heated fluid
shell to develop. After sufficient time, the fluid profile reaches an
asymptotic form.

The asymptotic fluid profile takes a qualitatively different form
depending on the expanding-bubble asymptotic wall speed
$\xiw$\footnote{Note that we distinguish between the asymptotic wall
speed $\xiw$ reached at late times for isolated expanding bubbles, and
the wall velocity $\vw$ measured or observed at a given time.}. If the
wall speed is subsonic, the transition front propagates as a
deflagration. In a deflagration, the fluid is accelerated and heated
at a leading-edge shock front. The fluid reaches its peak velocity at
the transition boundary. When the fluid crosses the transition
boundary, it decelerates and ends up at rest inside the bubble.

Walls travelling faster than the Jouguet detonation speed $\cJ$ give
rise to a detonation. The Jouguet detonation speed is dependent on the
transition strength, but it is always larger than the speed of
sound. In such a transition, the bubble wall hits fluid that is at
rest.  The fluid is then heated and accelerated as it crosses the
transition boundary, before decelerating inside the bubble.

In deflagrations, the heated fluid precedes the advancing bubble
wall. In sufficiently strong phase transitions, when the heated fluid
shells of multiple bubbles meet, the temperature of the metastable
state can become substantially higher than the nucleation
temperature. This effect has been noted recently in hydrodynamical
simulations of multiple bubbles colliding in 3D, see
Ref.~\cite{Cutting:2019zws}. In that paper, it was found that these
heated regions of the metastable state could persist for long periods
of time, effectively extending the duration of the phase transition
and reducing the wall velocity. These long-lasting regions were
referred to as \textsl{droplets} of the metastable state, as these
have been discussed previously in the literature, in particular in
reference to the QCD phase transition
\cite{Kajantie:1986hq,Rezzolla:1995kv,Rezzolla:1995br}. It should be
noted that droplets cannot form for detonations as the fluid is at
rest ahead of the bubble wall.

\begin{figure*}[htbp]
  \centering
  \subfigure[\,Early]{\includegraphics[width=0.32\textwidth]{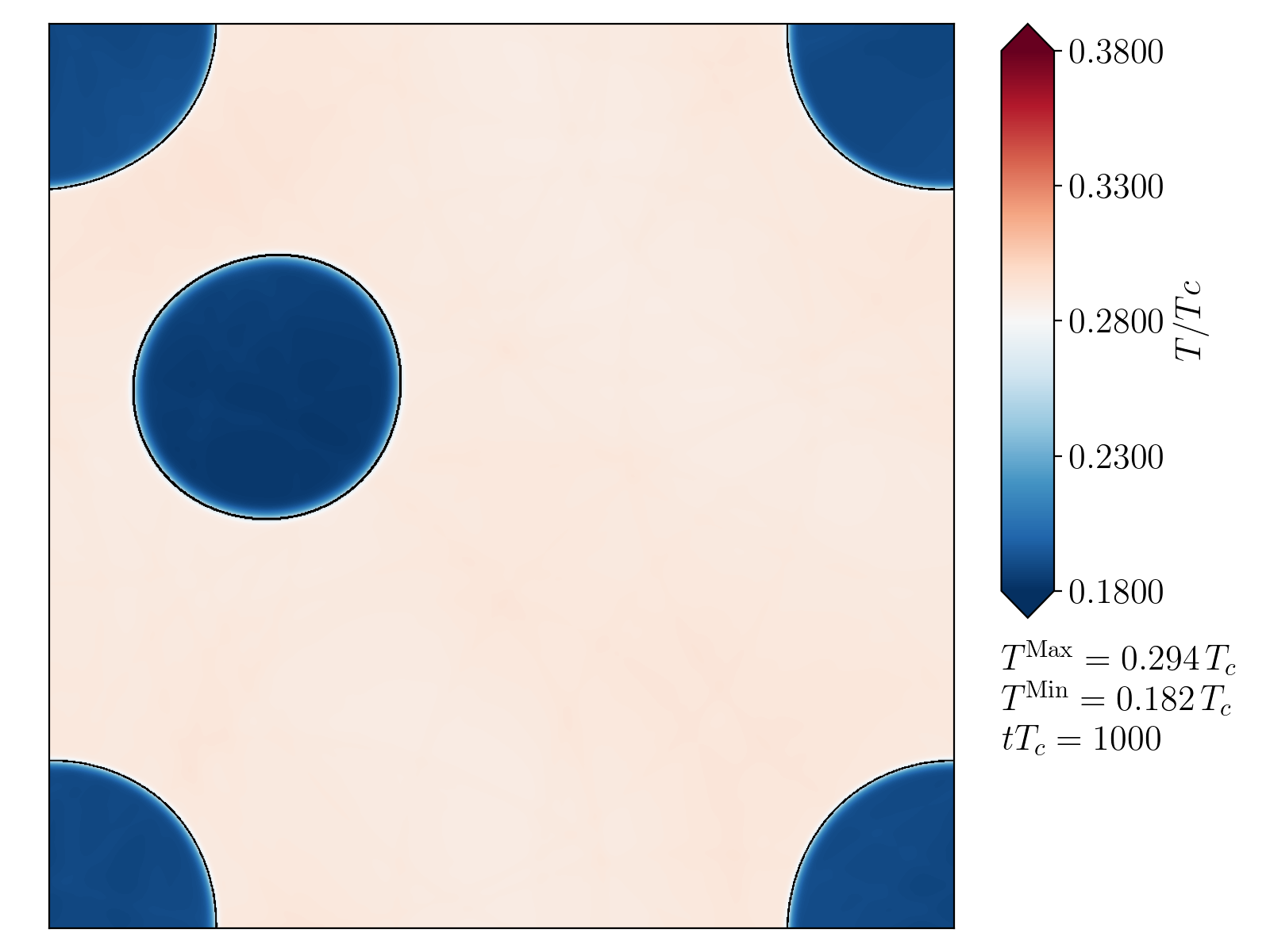}}
  \subfigure[\,Intermediate]{\includegraphics[width=0.32\textwidth]{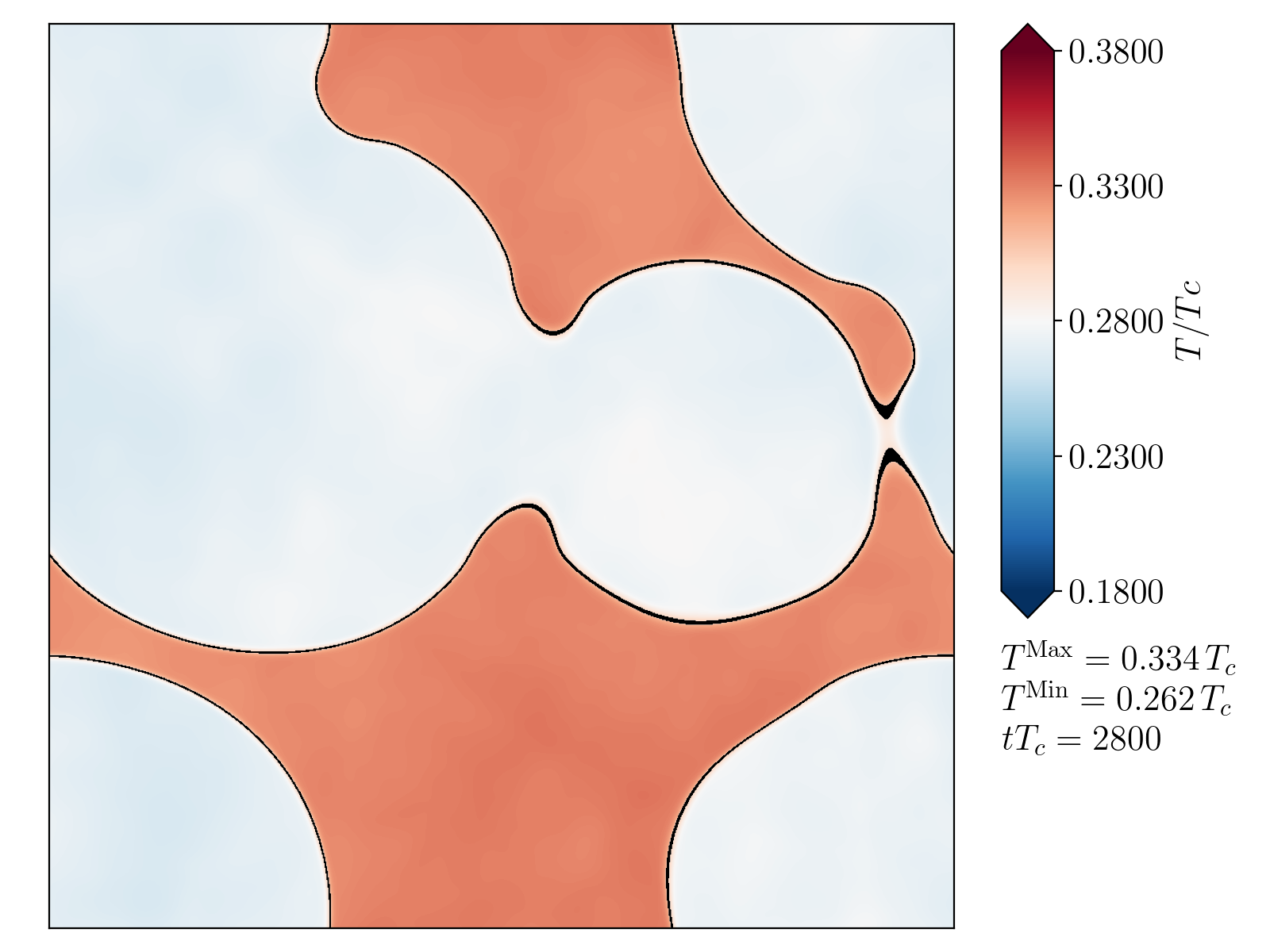}}
  \subfigure[\,Late]{\includegraphics[width=0.32\textwidth]{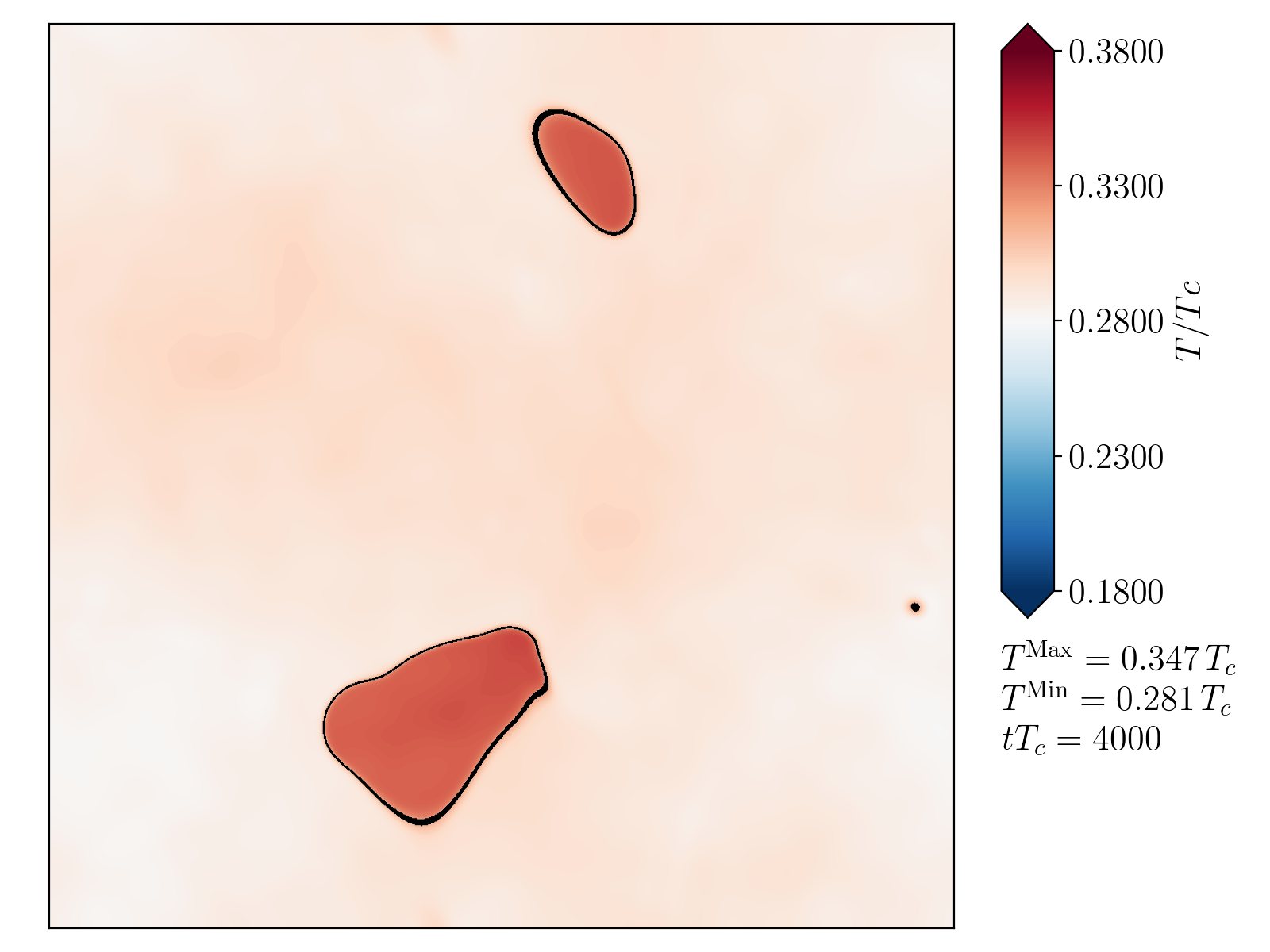}}\\
  \caption{2D slices of the temperature during a 3D simulation from
    Ref.~\cite{Cutting:2019zws}, showing eight bubbles expanding with
    an asymptotic expanding-bubble wall speed $\xiw=0.24$ and a
    transition strength $\Str=0.34$. The simulation had periodic
    boundary conditions and one of the bubbles was nucleated in the
    corners of the slice. The nucleation temperature was $\TN= 0.28 \,
    \Tc$. Black lines are used to highlight the phase transition
    boundary.}\label{fig:TSclices}
\end{figure*}

In Fig.~\ref{fig:TSclices}, we show an example of the evolution of the
temperature for a deflagration (the results of this simulation were
previously presented in Ref.~\cite{Cutting:2019zws}).  Three snapshots
of the temperature are shown, taking 2D slices through the 3D
simulation, first early on in the phase transition while the bubbles
remain isolated, then at an intermediate time where the heated fluid
shells have collided, and finally at a late time where only heated
droplets of the metastable state remain. In the simulation shown,
$\eta$ was chosen such that the asymptotic expanding-bubble wall speed
of isolated bubbles was $\xiw=0.24$, and the transition strength was
$\Str=0.34$. As can be seen, once the fluid shells collide, the phase
boundary begins to deform. A clear temperature difference is seen
between the metastable and true vacuum regions. At late times only
small droplets of the metastable state persist.

\begin{figure}[htbp]
  \centering
\includegraphics[width=0.48\textwidth]{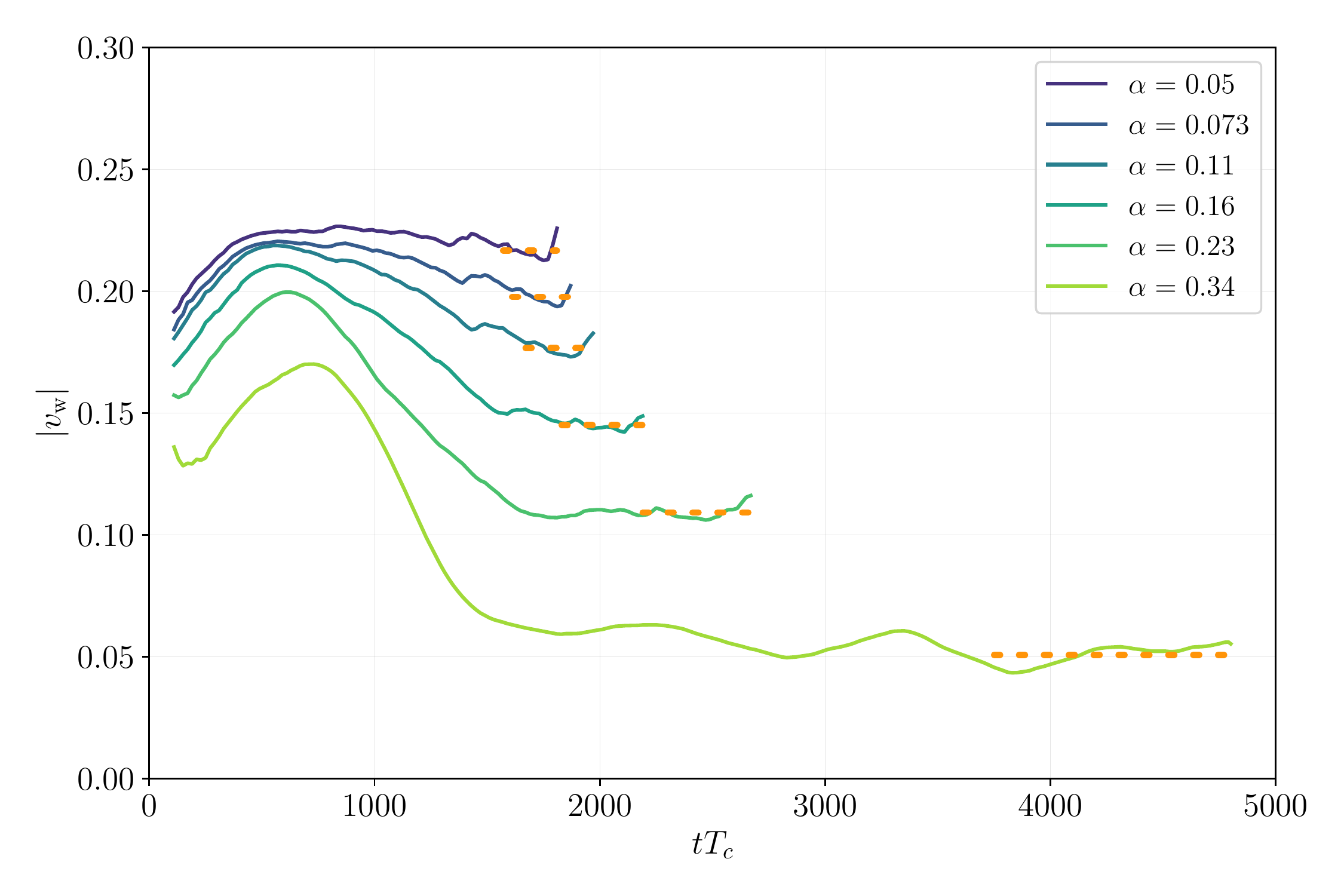}
\caption{Wall speed in 3D simulations from Ref.~\cite{Cutting:2019zws}
  of colliding bubbles with asymptotic expanding-bubble wall speeds of
  $\xiw=0.24$.  The different colours correspond to simulations with
  different transition strengths. The lines end when the metastable
  state takes up less than $2\%$ of the total volume. A late-time wall
  speed for each simulation is shown with an orange dashed line. The
  line extends over the times for which $\vws$ is fitted,
  corresponding to when the broken phase takes up between $10\%$ and
  $2\%$ of the total volume. See Table~\ref{tab:results} for the
  late-time values of $\vws$. See Appendix~\ref{app:wallspeed} for a
  discussion of wall speed estimators.}\label{fig:3DWallSpeed}
\end{figure}

The results of Ref.~\cite{Cutting:2019zws} indicated that that the
walls slowed down as the transition approached completion, and that
the effect was more pronounced for stronger transitions.  To
illustrate how the wall speed slows as droplets form, we have taken
simulations from Ref.~\cite{Cutting:2019zws} with the same asymptotic
expanding-bubble wall speed $\xiw=0.24$ and a variety of transition
strengths $\Str$. For each simulation, we plot the evolution of the
wall speed $\vw$ as a function of time in
Fig.~\ref{fig:3DWallSpeed}. As can be seen, initially the walls
accelerate towards $\xiw$, but at some point the fluid shells begin to
collide and thus heat up the metastable state.  At the same time the
phase boundaries start to decelerate. The stronger the phase
transition, the more the metastable state is heated above the
nucleation temperature, and the more the phase boundary decelerates.

The formation of droplets was associated with a decrease in the
kinetic energy production and substantial suppression of the
gravitational wave signal. The deceleration of the phase boundary
could also have an effect on baryogenesis, as it has been shown that
the efficiency of generating a baryon asymmetry has a strong wall
velocity dependence (see e.g.
Refs.~\cite{Cline:2021iff,Dorsch:2021ubz}).

\subsection{Similarity solutions}
\label{sec:similarity}

When the phase boundary of a droplet or a bubble reaches a terminal
wall velocity, the fluid profile approaches an asymptotic shape. To
find the form that this asymptotic profile takes, we need to match the
fluid velocity and enthalpy across the phase boundary.

Taking the energy--momentum tensor for a perfect fluid,
\begin{equation}
  T^{\mu\nu}= (\epsilon + p)U^\mu U^\nu + g^{\mu\nu}p\text,
\end{equation}
with enthalpy $w=\epsilon + p$, the conservation of energy and
momentum density across the phase boundary leads to
\begin{align}
  w_+ \tilde\gamma_+^2 \tilde v_+^2+p_+ &= w_- \tilde\gamma_-^2 \tilde v_-^2+p_-\text, \label{eq:sim-match1} \\
  w_+ \tilde \gamma_+^2 \tilde v^2_+&= w_- \tilde \gamma_-^2 \tilde v^2_- \text, \label{eq:sim-match2}
\end{align}
where $\tilde v$ and $\tilde \gamma$ refer to the fluid velocity and
the corresponding Lorentz factor in a frame moving along with the
phase boundary. Subscripts $+$ and $-$ refer to the quantities just
ahead and just behind the wall, respectively.

Once the velocity and enthalpy have been obtained on both sides of the
phase boundary, it is possible to find a solution for the rest of the
fluid profile by considering the continuity equation,
\begin{equation}
 \partial_\nu T^{\mu \nu}=0\text.
\end{equation}
Imposing spherical symmetry, it is possible to find differential
equations for $w$ and $v$ in terms of a self-similarity variable
constructed from a combination of a radial and time coordinate, $\xi=
r/\ts$~ \cite{Steinhardt:1981ct,KurkiSuonio:1984ba}. These
differential equations can be written in the following parametric
form, see e.g. Refs.~\cite{Espinosa:2010hh,Hindmarsh:2019phv}:
\begin{align}
  \frac{\mathrm{d}\xi}{\mathrm{d}\uptau} &= \xi \left[ (\xi - v)^2 - \cs^2(1-\xi v)^2\right]\text, \label{eq:sim-xi} \\
  \frac{\mathrm{d}v}{\mathrm{d}\uptau} &=  2v\cs^2 (1-v^2)(1-\xi v)\text, \label{eq:sim-v} \\ 
  \frac{\mathrm{d}w}{\mathrm{d}\uptau} &= w\left(1+\frac{1}{\cs^2}\right) \gamma^2 \mu \frac{\mathrm{d}v}{\mathrm{d}\uptau} \label{eq:sim-w}\text,
\end{align}
where
\begin{equation}
  \mu(\xi,v) = \frac{\xi - v}{1-\xi v}
\end{equation}
is the fluid velocity at $\xi$ in a frame moving with velocity $\xi$.

\begin{figure}[htbp]
  \centering
\subfigure[]{\includegraphics[width=0.48\textwidth]{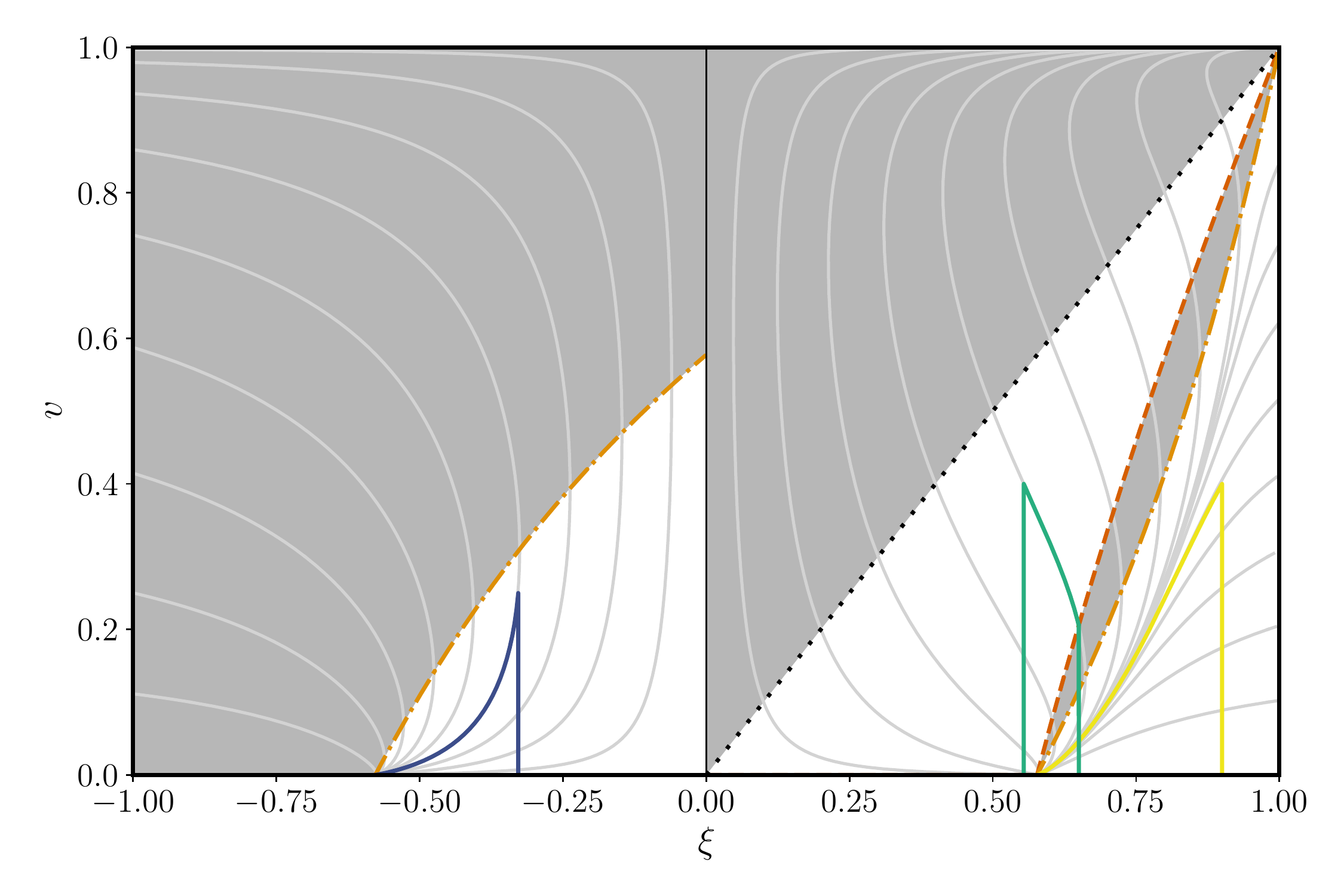}}
\caption{ Similarity curves for the fluid velocity profiles obtained
  from solving Eqs.~(\ref{eq:sim-xi}) and (\ref{eq:sim-v}). Positive
  values of $\xi$ correspond to bubbles and negative values to
  droplets. The similarity curves are shown in light grey, and the
  solid-coloured lines show example profiles, from left to right, for
  a droplet, a deflagration and a detonation. The two orange
  dash-dotted lines indicate the sound speed in a frame moving at
  velocity $\xi$, the red dashed line indicates the velocity of a
  deflagration shock front, and the black dotted line indicates the
  maximum fluid velocity allowed for a bubble profile. Regions with
  unphysical velocity profiles are shaded dark grey. For a thorough
  treatment on similarity curves in a cosmological transition, see
  Chapter 11.1 in
  Ref.~\cite{RezzollaZanotti}.\label{fig:SimilarityCurves}}
\end{figure}

\begin{figure}
  \centering
  \includegraphics[width=0.48\textwidth]{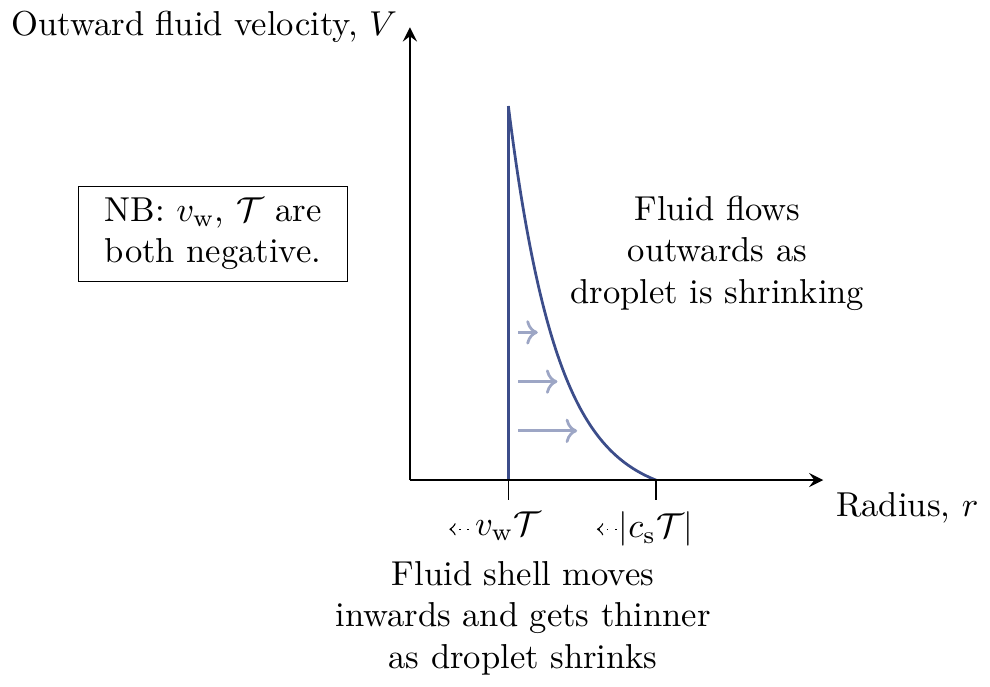}
  \caption{Sketch of a droplet similarity solution in context. The
    fluid flows outwards as the phase boundary moves inwards (at speed
    $\vws$). The wedge of fluid becomes thinner with time, as the
    outer boundary of the wedge also travels inwards (at speed
    $\cs$). However, the width of the wedge in terms of the parameter
    $\xi = r/\mathcal{T}$ remains constant.\label{fig:DropletSketch}}
\end{figure}

Solving these differential equations gives us a set of
similarity-profile curves, with the solutions for the velocity
profiles shown in Fig.~\ref{fig:SimilarityCurves}.  Values of $\xi>0$
correspond to expanding bubbles. In this case, $\ts$ is the time since
the nucleation of the bubble, and $r$ the distance from the bubble
centre. On the other hand, profiles with $\xi<0$ correspond to
shrinking droplets. Then, $r$ is the distance from the droplet centre,
and $\ts \Tc=0$ corresponds to the time at which the droplet
evaporates, with the droplet radius $\Rd\rightarrow +\infty$ when
$\ts\rightarrow -\infty$. From the plots, it can be seen that the
velocity curves have fixed points for $\uptau$ at $(\xi,v)=(\pm
\cs,0)$. For an indicative sketch of the droplet case, see
Fig.~\ref{fig:DropletSketch}.

In order to construct the self-similar profile, a wall speed and a
peak fluid velocity are chosen, which sets the initial point on the
$(\xi\text, v)$ plane. Using Eqs.~(\ref{eq:sim-xi})
and~(\ref{eq:sim-v}), the velocity profile can be constructed by
integrating backwards towards the fixed point at
$(\xi\text,v)=(\cs\text,0)$. Also drawn on the plot are lines
indicating the sound speed in a frame moving at velocity $\xi$, i.e.
$v=\mu(\xi,\pm \cs)$. For bubbles, this sets the maximum fluid
velocity for a detonation, whereas for droplets it separates the
physical subsonic deflagration profiles from the unphysical sonic and
supersonic deflagrations (i.e. droplets with $\xiw \geq \cs$). For
bubbles, the fluid velocity cannot exceed $v=\xi$, as in the wall
frame this would mean that the fluid would be flowing outwards from
inside the bubble. Finally, the speed at which the leading edge of a
deflagration shock propagates is shown for bubbles. For deflagrations,
when the velocity similarity curve hits this line, there is a
discontinuity in the profile with $v=0$ for larger values of
$\xi$. Similar curves can be constructed for the enthalpy profiles.

\subsection{Kinetic energy fraction and gravitational waves}

An important quantity for the generation of gravitational waves is the
kinetic energy fraction,
\begin{equation}
  K = \frac{\langle w\gamma^2 v^2 \rangle}{\langle w\gamma^2 v^2 + \epsilon \rangle} \equiv \frac{\langle w\gamma^2 v^2 \rangle}{\langle e \rangle}  \text,
  \label{eq:KDefn}
\end{equation}
namely, the volume-averaged kinetic energy relative to the
volume-averaged total energy density $\langle e \rangle$ at the end of
the transition. It has previously been shown that the amplitude of the
gravitational wave power spectrum is proportional to
$K^2$~\cite{Hindmarsh:2015qta,Hindmarsh:2017gnf}. Estimates of $K$
that do not require expensive 3D numerical simulations are therefore
highly desirable. The typical procedure is to find the asymptotic
profile of a bubble expanding in isolation and to assume that the
value of $K$ extracted from the fluid profile matches well onto the
final result obtained from many bubbles colliding in the 3D numerical
simulation. While this procedure is known to be accurate for weak and
intermediate-strength
transitions~\cite{Hindmarsh:2015qta,Hindmarsh:2017gnf}, it breaks down
for strong transitions where heated fluid shells start interacting
with each other nonlinearly. In particular, heated droplets of the
metastable state form in strong deflagrations, and in
Ref.~\cite{Cutting:2019zws} it was proposed that the formation of
these droplets was associated with a drop in $K$ relative to that
predicted from an isolated bubble.

In this work, we track the kinetic energy fraction in the fluid for
spherically symmetric bubbles and droplets. We define the kinetic
energy fraction of a bubble as
\begin{equation}
  \Kb(t) = \frac{\int^{\infty}_0 \mathrm{d}r\, r^2 w\gamma^2 v^2}{\int^{\Rb(t)}_0 \mathrm{d}r\, r^2 e(t=0) }\text,
  \label{eq:KbDefn}
\end{equation}
with $\Rb(t)$ the bubble radius at time $t$ after nucleation. The
numerator is the kinetic energy of the fluid shell around the
bubble. For the denominator, we consider the region converted into the
true vacuum by the bubble at time $t$, and find the total energy of
the system that was contained within that volume at $t=0$. We
sometimes use the notation $\Kbref$, which indicates the value
$\Kb(t)$ at the time $t$ for which the radius of the bubble is given
by a reference radius, $\Rb(t)=\Rbref$. This is a unique time, as the
bubble radius is monotonically increasing with respect to $t$ for an
isolated expanding bubble.

Similarly, we define the kinetic energy fraction for a droplet to be
\begin{equation}
  \Kd(t) = \frac{\int^{\infty}_0 \mathrm{d}r\, r^2 w\gamma^2 v^2}{\int^{\RdI}_{\Rd(t)} \mathrm{d}r\, r^2 e(t=0) }\text,
  \label{eq:KdDefn}
\end{equation}
with $\RdI$ the initial droplet radius. Analogous to the previous
expression, for the denominator we consider the region converted into
the true vacuum at time $t$ and calculate the initial total energy of
the system contained in that region. Note the different limits of
integration when compared with Eq.~(\ref{eq:KbDefn}).

Comparing these two quantities for a droplet released from $\RdI$ and
a bubble for which $\Rb(t)=\RdI$ gives a measure of the relative
efficiency of producing kinetic energy for droplets.

Another quantity that is often measured in simulations is the
enthalpy-weighted mean square fluid 4-velocity
\begin{equation}
  \label{eq:UbarfDefn}
  \Ubarf^2 = \frac{\langle w \gamma^2 v^2 \rangle}{ \overline w}\text.
\end{equation}
with $\overline{w}$ the mean enthalpy density. $\Ubarf^2$ is related
to the kinetic energy faction via
\begin{equation}
  K = \Gamma \Ubarf^2\text,
\end{equation}
where
\begin{equation}
  \Gamma = \frac{\overline{w}}{\overline{e}}
\end{equation}
is the mean adiabatic index in the fluid.

While we could in principle also estimate $\Ubarf^2$ from our
spherical simulations, in practice this leads to severe
inconsistencies for strong transitions where $w$ after the transition
can differ significantly from its initial value.  To illustrate this,
let us define
\begin{equation}
  \Ubarfb^2(t) = \frac{\int^\infty_0 \mathrm{d}r\, r^2 w \gamma^2 v^2 }{\int^{\Rb(t)}_0 \mathrm{d}r\, r^2 w_\mathrm{n}},
  \label{eq:UbarfbDefn}
\end{equation}
for bubbles in line with previous three-dimensional simulation work.

The denominator for $\Kb(t)$ in Eq.~(\ref{eq:KbDefn}) is the initial
total energy of the system for the volume converted into the true
vacuum by a bubble at time $t$. For $\Ubarfb^2(t)$ it is the initial
enthalpy of the system for the equivalent volume. However, while the
energy is conserved, the enthalpy is not. Therefore, while the
denominator of Eq.~(\ref{eq:KbDefn}) is an accurate estimate of the
total energy density in Eq.~(\ref{eq:KDefn}), we cannot easily
estimate the denominator of Eq.~(\ref{eq:UbarfDefn}) for an isolated
bubble or droplet.

\section{Methods}
\label{sec:compmethods}

In this study, we present results from evolving spherically symmetric
bubbles and droplets in the coupled field--fluid model. Due to
spherical symmetry, we are able to evolve the equations of motion of
Eqs.~(\ref{eq:phi-EOM}-\ref{eq:Z-EOM}) on a 1D lattice. To do this, we
use a simplified 1D version of the SCOTTS code used in
Refs.\cite{Hindmarsh:2015qta,Hindmarsh:2017gnf,Cutting:2019zws}. This
is based on the 1D code of Ref.~\cite{Ignatius:1993qn}, which
implements a Minkowski space version of the algorithms outlined in
Refs.~\cite{1983ApJ...273..428C,1984ApJS...54..229C,Kurki-Suonio:1987mrt}. For
the evolution of the scalar field, a Crank--Nicolson
update~\cite{crank_nicolson_1947} is used. For the hydrodynamical
evolution, the code uses operator splitting to update each term in
Eqs.(\ref{eq:E-EOM}-\ref{eq:Z-EOM}) and upwind donor cell for the
advection terms.

Our initial conditions in the spherical simulations depend on whether
we are performing a simulation of a droplet or a bubble. For bubbles,
we initially prepare the scalar field to be in the broken phase close
to the origin and in the symmetric phase away from the centre of the
bubble. We use a Gaussian profile
\begin{equation} \label{eq:bubble-initial}
\phi(r) = \phib \, \mathrm{exp}\left( \frac{-r^2}{2 \Rc^2} \right)
\end{equation}
with
\begin{equation} \label{eq:critical-radius}
  \Rc = \frac{2 \sigma}{\potT(0,\TN)-\potT(\phib,\TN)},
\end{equation}
the critical radius in the thin-wall approximation.  In the equation
above, $\sigma$ refers to the surface tension in the thin-wall
approximation and is given by
\begin{equation}
  \sigma = \frac{{\left(\mu^2 + \mu\sqrt{\mu^2 - 4 M^2 \lambda} -2 M^2 \lambda \right)}^{3/2} } {24 \lambda^{5/2}}\text.
\end{equation}
    
For droplets, we instead prepare the field profile to be in the
symmetric phase at the origin and in the broken phase far away from
it. We choose an initial droplet radius, $\RdI$, and then fix the wall
profile using a $\mathrm{tanh}$-shape. The scalar field is then set
using
\begin{equation} \label{eq:droplet-initial}
\phi(r) = \frac{\phib}{2} \left(1 - \mathrm{tanh}\left(\frac{\RdI - r}{\lw}\right)\right)\text,
\end{equation}
with wall thickness $\lw$ given by the thin-wall approximation,
\begin{equation}
  \label{eq:wall_thickness}
  \lw = \frac{2}{\sqrt{\potT''(\phib,\Tc)}}\text.
\end{equation}
Note that, even for bubbles nucleated with a Gaussian or `thick-wall'
profile like that in Eq.~\ref{eq:bubble-initial}, we still expect the
phase boundary to relax to a $\mathrm{tanh}$-like profile as it
expands. This is due to the phenomenological friction term of
Eq.~(\ref{eq:coupling}), and motivates the form of the initial droplet
profile given above.

In the initial conditions for both droplets and bubbles, we initialise
the fluid such that $T=\TN$ and $\mathbf{v}=0$ everywhere. Note that
this differs from other studies of the evolution of droplets, where
the simulations are initialised with an initial temperature jump
across the phase
boundary~\cite{Rezzolla:1995kv,Kurki-Suonio:1995yaf,Rezzolla:1995br}. In
this work we are also interested in the heating caused by the collapse
of the droplet and the effect this has on the phase boundary velocity
and kinetic energy production. We hence leave simulations in which a
temperature jump is initialised based on values extracted from
droplets formed in 3D multi-bubble simulations for future work.

We match our potential parameters to those used in
Ref.~\cite{Cutting:2019zws} for ease of comparison with earlier 3D
multi-bubble simulations.  The potential parameters and the
corresponding wall thickness are given in Table~\ref{tab:pot-params}
(see Appendix.~\ref{app:fractionalchange} for an investigation of
varying the fractional change in the number of degrees of freedom).

We use a lattice spacing $\Delta x\, \Tc = 1.0$ in all the
simulations. For our droplet simulations, we set the initial
  droplet radius $\RdI\, \Tc= 2\,\times\, 10^{4}$ (see
  Appendix~\ref{app:convergence} for an investigation of the
  consequences of varying the droplet radius).  Other simulation
parameters, such as the timestep $\Delta t$ and the simulation
duration $t_\mathrm{fin}$, are given for the 1D runs in
Table~\ref{tab:init-params}.  We set the number of simulation sites $L
= t_{\mathrm{fin}}/\Delta x$.

We perform simulations of bubbles which expand with an asymptotic wall
speed $\xiw=0.24$. In this model, the strength $\alpha$ is varied
  by varying $\TN$, keeping all other input parameters constant. We
pick a range of $\TN$ that gives transitions spanning from $\Str
=0.005$ up to $\Str=0.41$, which is close to the maximum $\Str$
allowed\footnote{As discussed at the end of
Section~\ref{sec:similarity}, the peak fluid velocity in a
self-similar flow cannot exceed the wall velocity. As increasing
$\Str$ while keeping $\xiw$ fixed increases the peak fluid velocity,
this implies there is a maximum value of $\Str$ for a self-similar
flow with a given $\xiw$.}  for $\xiw=0.24$. For each value of $\TN$,
we find the value of $\eta$ that gives $\xiw=0.24$ for a bubble, and
then also run an equivalent droplet simulation with this value of
$\eta$. We list these values in Table~\ref{tab:init-params}.  We also
perform comparisons with 3D multi-bubble simulations. These correspond
to the simulations with $\xiw=0.24$ in Ref.~\cite{Cutting:2019zws},
which we list again here in Table~\ref{tab:3D-simulations} for
convenience.

\begin{table}
  \centering
   \begin{tabularx}{\columnwidth}
     {XXXXXXX}
     
   \toprule \toprule 
   \\ [-0.5em]
      $g_*$ & $M^2/\Tc^{2}$ & $\mu/\Tc$ & $\lambda$ & $\phib/\Tc$ & $\Delta a/a_0$ & $\lw \,\Tc$ 
   \\ [0.5em] \midrule \\
      106.75 & 0.0427 &  0.168 & 0.0732 & 2.00 & 0.0059 & 5.23 
   \\ [0.5em] \bottomrule \bottomrule
	\end{tabularx}  
        \caption{Table of key constant quantities for the computations
          in this paper. The input parameters are the effective
          degrees of freedom during the transition $g_*$ and the
          potential parameters $M^2$, $\mu$ and $\lambda$. Using the
          potential parameters, the broken phase scalar field minimum
          $\phib$, the change in degrees of freedom $\Delta a$ during
          the transition, Eq.~(\ref{eq:delta_a}), and the reaction
          front (wall) thickness $\lw$, Eq.~(\ref{eq:wall_thickness}),
          can be derived.\label{tab:pot-params}}
\end{table}

\begin{table*}[t]
\begin{tabularx}{2\columnwidth}
{ X@{\hspace{0.8cm}} X*{5}{d{-2}} @{\hspace{-0.8cm}}}
\toprule \toprule \\ [-0.5em]
\multicolumn{1}{X}{ Type} & 
\multicolumn{1}{X}{\hspace{-2mm} $\Str$} & 
\multicolumn{1}{X}{\centering $\hspace{4.4mm} \TN / \Tc$} & 
\multicolumn{1}{X}{\centering $\hspace{-0.4mm} \eta / \Tc$} &
\multicolumn{1}{X}{\centering $\hspace{4mm} \Delta t \, \Tc$} & 
\multicolumn{1}{X}{\centering $\hspace{5mm} t_{\mathrm{fin}} \, \Tc$}  & 
\multicolumn{1}{X}{\centering $\hspace{9mm} \Rc \, \Tc$}
\\ [0.5em] \midrule \\
\multirow{8}{*}{Droplet}
& 0.0050
& 0.79 & 0.68
& 0.2 &  3.0 \times 10^5 & \text{---} 
\\
& 0.050
& 0.45 & 1.2
& 0.2 & 3.0 \times 10^5 & \text{---} 
\\
& 0.073 
& 0.41 & 1.3
& 0.2 & 3.0 \times 10^5 & \text{---}
\\
& 0.11 
& 0.37 & 1.5
& 0.2 & 3.0 \times 10^5 & \text{---}
\\
& 0.16 
& 0.33 & 1.8
& 0.2 & 3.0 \times 10^5 & \text{---}
\\
& 0.23 
& 0.30 & 2.4
& 0.2 & 3.0 \times 10^5 & \text{---}
\\
& 0.34 
& 0.28 & 5.1
& 0.2 & 6.4 \times 10^5 & \text{---}
\\
& 0.41 
& 0.26 & 11
& 0.1 & 6.4 \times 10^5 & \text{---}
\\ [0.5em] 
\midrule \\ [-0.5em] 
\multirow{8}{*}{Bubble}
& 0.0050
& 0.79 & 0.68
& 0.2 & 0.9 \times 10^5 & 12
\\
& 0.050
& 0.45 & 1.2
& 0.2 & 0.9 \times 10^5 & 7.7
\\
& 0.073 
& 0.41 & 1.3
& 0.2 & 0.9 \times 10^5 & 7.6
\\
& 0.11 
& 0.37 & 1.5
& 0.2 & 0.9 \times 10^5 & 7.5
\\
& 0.16 
& 0.33 & 1.8
& 0.2 & 0.9 \times 10^5 & 7.5
\\
& 0.23 
& 0.30 & 2.4
& 0.2 & 0.9 \times 10^5 & 7.4
\\
& 0.34 
& 0.28 & 5.1
& 0.2 & 0.9 \times 10^5 & 7.4
\\
& 0.41 
& 0.26 & 11
& 0.1 & 0.9 \times 10^5 & 7.4
\\ [0.5em] \bottomrule \bottomrule
\end{tabularx}
\caption{Parameters used for the spherically symmetric 1D runs, split
  according to droplet (Eq.~(\ref{eq:droplet-initial})) or bubble
  (Eq.~(\ref{eq:bubble-initial})) initial condition type. For each
  transition strength $\Str$, we list the nucleation temperature $\TN$
  relative to the critical temperature $\Tc$, the friction parameter
  $\eta$, the timestep $\Delta t$ and the final time in the simulation
  $t_\mathrm{fin}$. For the bubble runs we also list the critical
  radius $\Rc$ as given by
  Eq.~(\ref{eq:critical-radius}). \label{tab:init-params}}
\end{table*}

\begin{table}[t]
  \begin{tabularx}{1\columnwidth}{X X X X@{\hspace{-1.2cm}}}
\toprule \toprule \\ [-0.5em]
 $\Str$ & 
$\TN / \Tc$ & $ \eta / \Tc$ &  $ \Rc \, \Tc$ 
\\ [0.5em] \midrule  \\
0.050
& 0.45 & 1.2 & 7.7
\\
0.073 & 0.41 & 1.3  & 7.6
\\
0.11 & 0.37 & 1.5  & 7.5
\\
0.16 & 0.33 & 1.8  & 7.5
\\
0.23 & 0.30 & 2.4  & 7.4
\\
0.34 & 0.28 & 5.1  & 7.4
\\ [0.5em] \bottomrule \bottomrule
\end{tabularx}
\caption{List of 3D multi-bubble simulations from
  Ref.~\cite{Cutting:2019zws} used for comparison in this paper. In
  addition to the parameters given above, all simulations had
  $L^3=960^3$ lattice sites, $N_\mathrm{b}=8$ bubbles, $\Delta x \Tc =
  1.0$, $\Delta t \Tc = 0.2$ and final time $t_{\mathrm{fin}} \Tc =
  4.8 \times 10^3$.\label{tab:3D-simulations}}
\end{table}

\section{Results}\label{sec:Results}

In our spherical droplet simulations, the phase boundary is released
from rest with initial bubble radius $\RdI$ at the start of each
simulation. The droplet starts shrinking. At the same time, an
inward-moving fluid shell is generated in the interior of the droplet
and travels towards the centre of the droplet faster than the phase
boundary itself. As this fluid shell propagates, the temperature rises
and the phase boundary begins to slow. When the inward-moving fluid
shell hits the origin, it rebounds, and a reflected fluid shell is
driven back towards the phase boundary. When the reflected fluid shell
meets the phase boundary, part of the fluid shell can once more be
reflected towards the origin\footnote{Note that similar reflections of
the fluid at phase boundaries can be seen in the movies of the
temperature for the multi-bubble transitions in
Ref.~\cite{Cutting:2019zws}, which are available at
Ref.~\cite{external}.}. This process can happen several times before
the droplet evaporates.  The interaction of the fluid with the phase
boundary can significantly decelerate the phase boundary, and for
stronger transitions the droplet can even be forced to temporarily
grow in size before shrinking again.

\begin{figure*}[htbp]
    \centering
    \includegraphics[width=0.75\textwidth]{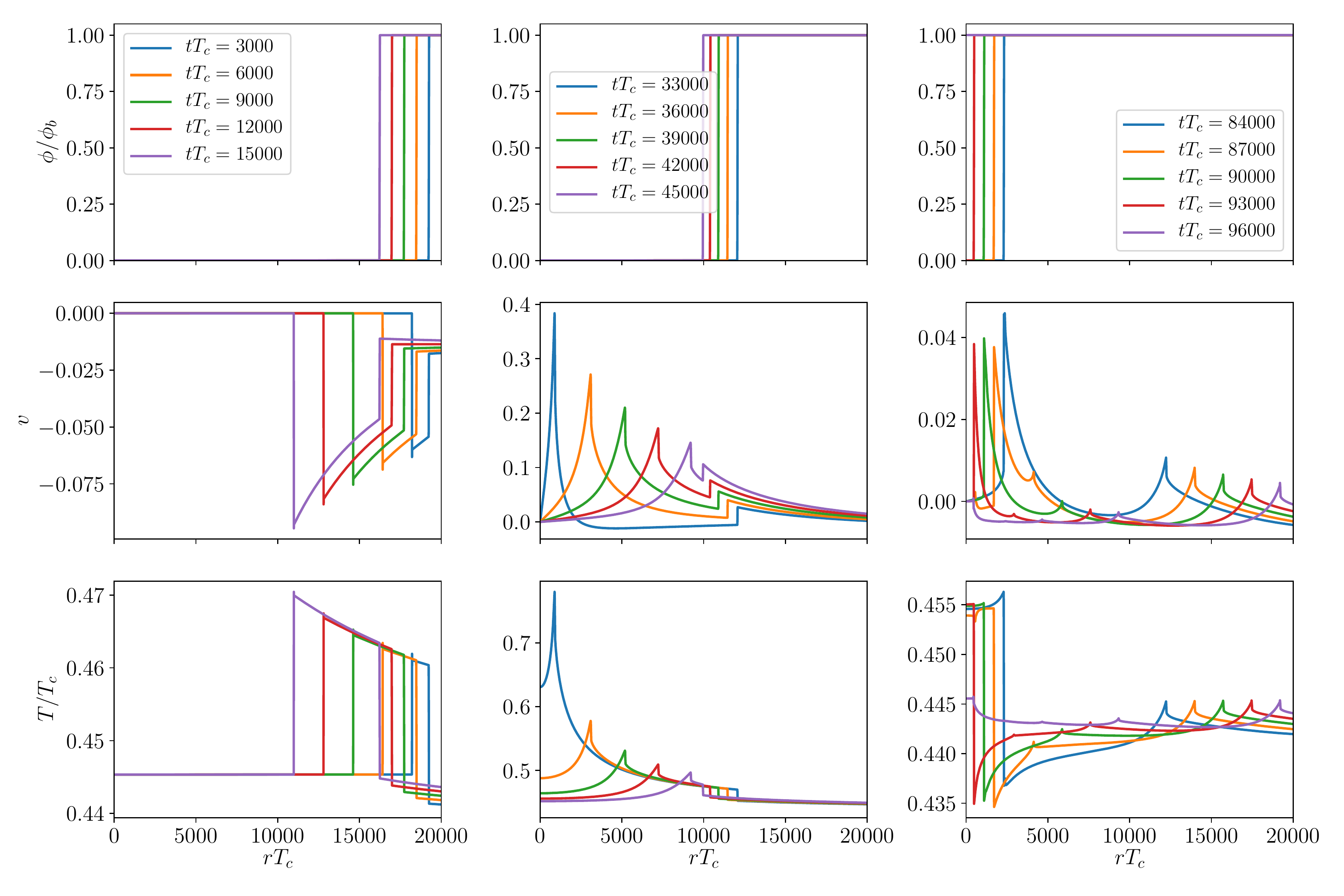}
    \caption{Evolution of $\phi$, $v$ and $T$ for a spherical droplet
      simulation with $\Str=0.05$. The top panel in each row shows the
      profile of the scalar field at various times, the middle panel
      shows the profile of the fluid velocity, and the bottom panel
      shows the temperature. The left column shows the evolution at
      early times, when the fluid shell is beginning to develop. The
      middle column shows intermediate times, after the fluid shell
      has been reflected at the origin. The right column shows late
      times, shortly before the droplet evaporates. As we do not find
      a similarity solution for weak transitions before the droplet
      evaporates, the late-time behaviour of the system appears less
      straightforward than for the strong transition seen below in
      Fig.~\ref{fig:lineoutalph0.34}. Note that the $y$-axis ranges
      change between columns for $v$ and $T$. A movie of this
      simulation is available at
      Ref.~\cite{cutting_daniel_2022_6390336} \label{fig:lineoutalph0.05}}
\end{figure*}

\begin{figure*}[htbp]
  \centering
  \includegraphics[width=0.75\textwidth]{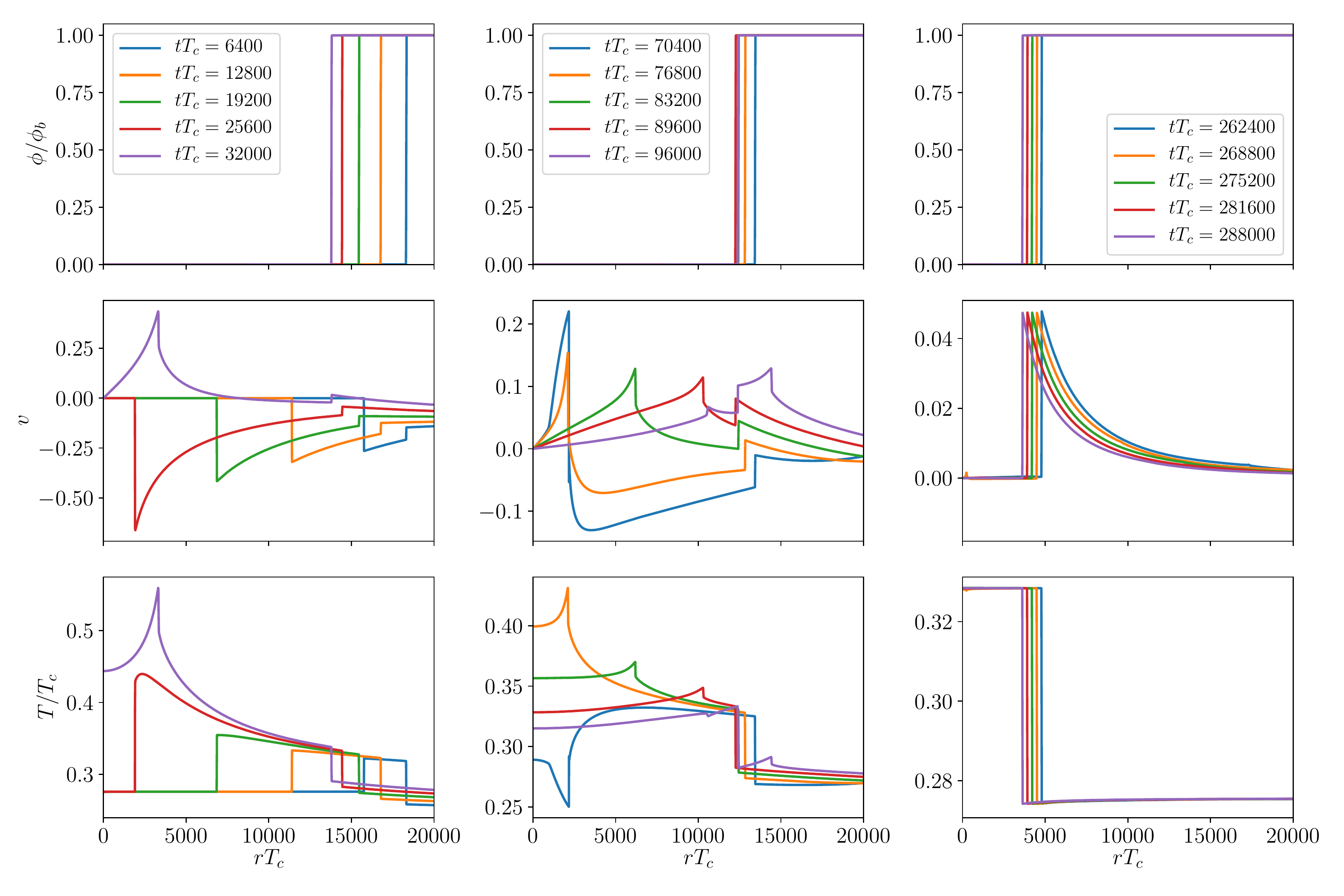}
  \caption{Evolution of $\phi$, $v$ and $T$ for a spherical droplet
    simulation with $\Str=0.34$. As in Fig.~\ref{fig:lineoutalph0.05},
    we show the situation at early, intermediate and late times. Note,
    however, that the exact times differ, as the droplet shrinks at a
    slower rate. Notably, the late-time behaviour in this case is that
    of a similarity solution of the fluid equations. Again, the
    $y$-axis ranges change between columns for $v$ and $T$. A movie of
    this simulation is available at
    Ref.~\cite{cutting_daniel_2022_6390336} \label{fig:lineoutalph0.34}}
\end{figure*}

We show the evolution of the scalar field, velocity and temperature
profiles for a droplet with a relatively weak transition strength
$\Str = 0.05$ in Fig.~\ref{fig:lineoutalph0.05}, and for one with a
stronger transition strength $\Str=0.34$ in
Fig.~\ref{fig:lineoutalph0.34}. The late-time evolution of the fluid
profile in the simulation of $\Str=0.34$ demonstrates self-similarity
once the interior of the droplet no longer contains any significant
fluid perturbations. Movies of these simulations are available at
Ref.~\cite{cutting_daniel_2022_6390336}.

\begin{figure}[htbp]
  \centering
  \includegraphics[width=0.48\textwidth]{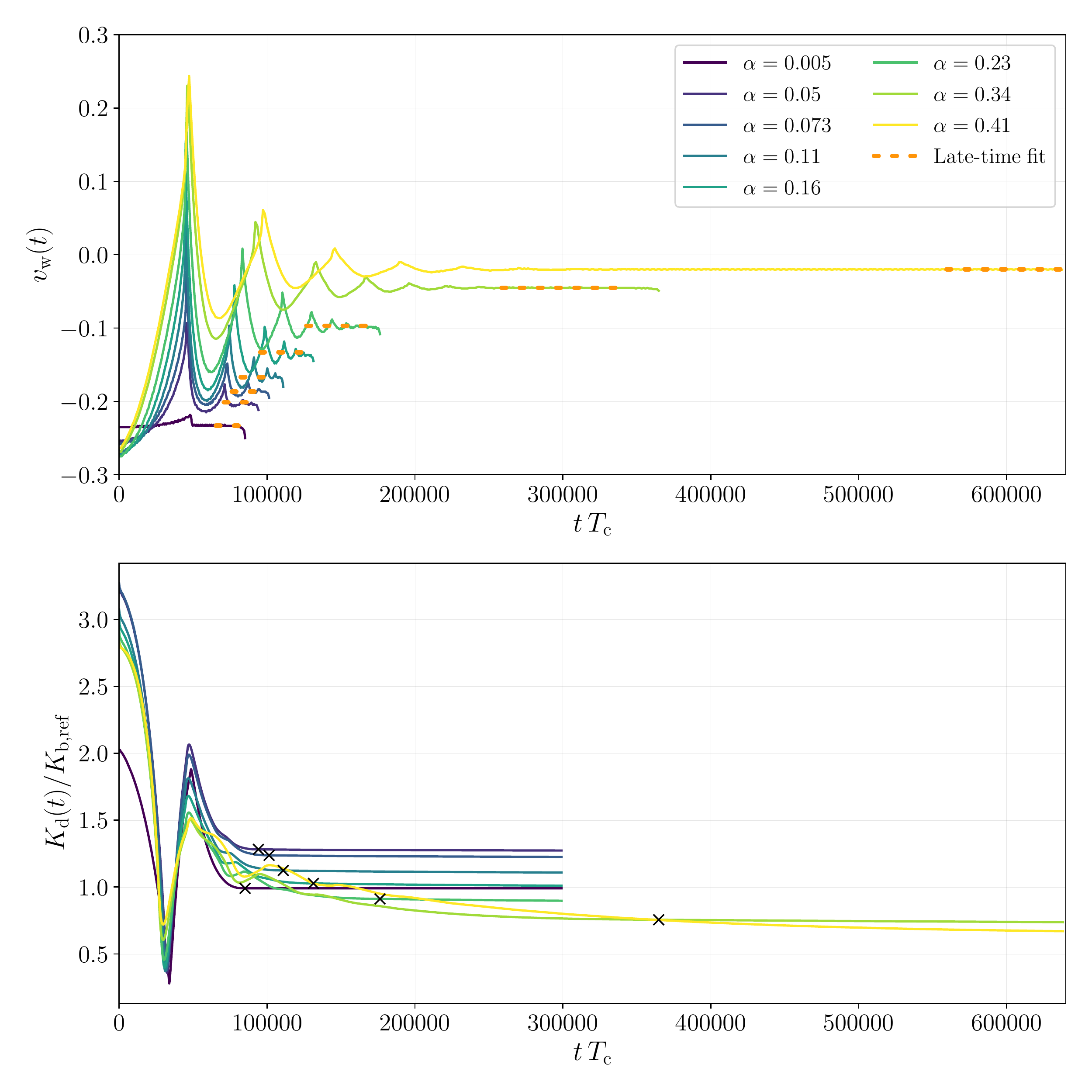}
  \caption{In the top panel we show the evolution of the wall velocity
    in the spherical droplet simulations. See
    Appendix~\ref{app:wallspeed} for a discussion on wall speed
    estimators.  Each solid line corresponds to an individual
    simulation with a given $\Str$. These lines end when the phase
    boundary reaches the origin. We average over $\vw$ when $0.05
    \RdI<\Rd<0.25 \RdI$ to find The values of $\vwass$ and the times
    averaged over are shown with dashed orange lines. In the bottom
    panel we show the evolution of $\Kd$ in each simulation. We
    normalise $\Kd$ (see Eq.~\ref{eq:KdDefn}) according to the value
    of $\Kbref$ (see Eq.~\ref{eq:KbDefn}) for a bubble with the same
    values of $\Str$ and $\eta$ and with radius $\Rbref=\RdI$. The
    black crosses on the bottom panel refer to the time at which the
    phase boundary reaches the origin. Values of $\vwass$ and of
    $\Kd/\Kbref$ at the black crosses are given in
    Table~\ref{tab:results}.\label{fig:rbtdot_K_wrt_time}}
\end{figure}

We track the velocity $\vw$ of the phase boundary in our spherical
droplet simulations (see Appendix~\ref{app:wallspeed}). We plot the
evolution of $\vw$ for each simulation in the upper panel of
Fig.~\ref{fig:rbtdot_K_wrt_time}. When $\vw$ is negative, the droplet
is shrinking, with the phase boundary moving towards the origin. For
positive values of $\vw$, the phase boundary is moving outwards,
causing the droplet to temporarily increase in size. For the smallest
transition strength, $\Str = 0.005$, we see that the phase boundary
travels towards the origin with speed close to that of an expanding
bubble, with $\vw \approx-0.24$. At larger transition strengths, $\vw$
rapidly decelerates as the phase boundary slows and the temperature
immediately inside the droplet increases. For the strongest
transitions it can be noticed that $\vw$ becomes temporarily positive.

As we noted above, the fluid shells can be reflected between the
origin and the phase boundary multiple times.  While this is
happening, the phase boundary velocity $\vw$ oscillates with shrinking
amplitude, before settling towards an asymptotic value. We can see
that the late-time wall velocity tends towards zero as the transition
strength increases. Furthermore, for the two strongest transitions,
$\Str=0.34$ and $\Str=0.41$, we see that the final wall velocity is
reached a long time before the droplet evaporates. Comparison with the
late-time evolution of the profiles in Fig.~\ref{fig:lineoutalph0.34}
indicates that a self-similar profile is obtained before the
evaporation of the droplet. We will discuss this in more detail in
Section~\ref{sec:self-similar}.

\begin{figure}[htbp]
  \centering
  \includegraphics[width=0.48\textwidth]{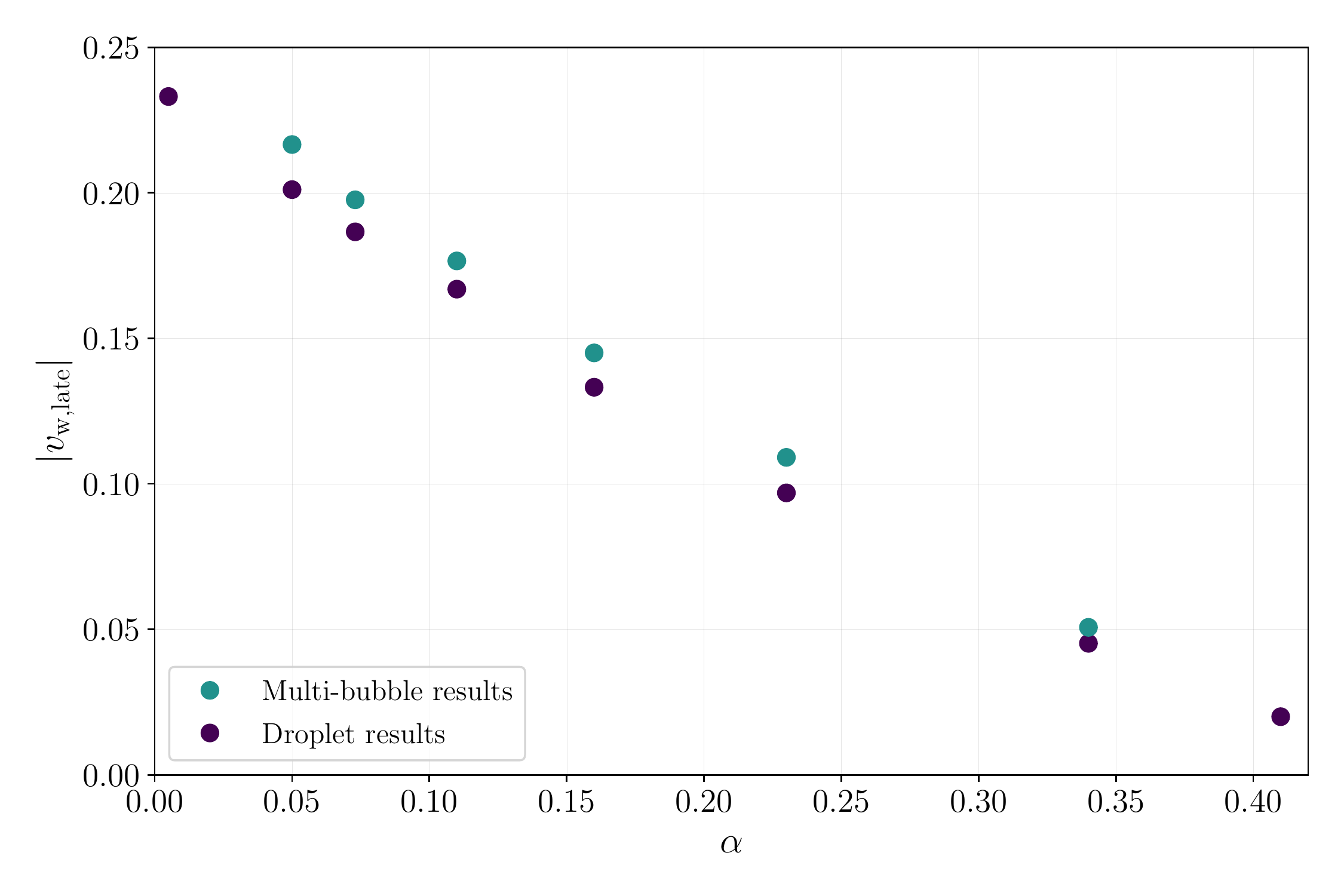}
  \caption{Late-time wall speeds $\left|\vwass\right|$ in both 3D
    multi-bubble simulations and spherical droplet simulations. Values
    are taken from Table~\ref{tab:results}, see the caption for
    details. }\label{fig:LateWallSpeeds}
\end{figure}

In Fig.~\ref{fig:LateWallSpeeds}, we compare the late-time wall
velocities $\vwass$ found in the spherical simulations with those
extracted from the multi-bubble simulations of
Ref.~\cite{Cutting:2019zws}. There is broad agreement across the range
of $\Str$ we consider, indicating that the late-time propagation of
the phase boundary in a droplet can be well modelled from a spherical
simulation. This seems independent of whether or not we reach a
similarity solution in the spherical droplet simulation.

Next, we look at the kinetic energy fraction $\Kd$ (see
Eq.~\ref{eq:KdDefn}) for droplets, noting that it is a key parameter
for predicting gravitational wave power spectra.  In the bottom panel
of Fig.~\ref{fig:rbtdot_K_wrt_time}, we show the evolution of $\Kd$ in
the aforementioned spherical droplet simulations.  We normalise $\Kd$
by using $\Kbref$ (see Eq.~\ref{eq:KbDefn}), which is taken from a
bubble with the same value of $\eta$ and $\Str$ at a given reference
radius $\Rbref$. In this case, we set $\Rbref=\RdI$. This quantity
then provides a measure of the relative efficiency for which a droplet
system produces kinetic energy compared to an isolated, expanding
bubble.

From Fig.~\ref{fig:rbtdot_K_wrt_time} we see that kinetic energy is
produced in excess of that expected from $\Kbref$. This is most likely
due to the initial shell of fluid produced by the droplet when it is
released from rest. Unlike an expanding bubble which grows to a large
size from a microscopic one, a droplet shrinks from a large size until
it evaporates. The effect of the initial conditions is therefore
emphasised in the droplet simulations. We see that $\Kd$ decreases
initially as the droplet shrinks, before rebounding when the fluid
shell hits the origin. After this, $\Kd$ slowly decreases until the
droplet evaporates. It is worth noting that if we used the similarity
solution as an initial condition for the fluid profile, then $\Kd$
would simply tend towards zero as the droplet would shrink, since the
fluid shell size of the similarity solution is relative to the radius
of the droplet.

However, it is not clear if a similarity solution would be reached for
a realistic three-dimensional simulation, where the initial droplet is
non-spherical and the initial velocity field is non-zero. Furthermore,
we need to know the final wall velocity for a given $\eta$
corresponding to the shrinking similarity solution in order to
generate appropriate initial conditions. Currently, this requires
running a simulation in any case, making the simulation of the true
shrinking similarity solution an iterative process.

\begin{figure}[htbp]
  \centering
  \includegraphics[width=0.48\textwidth]{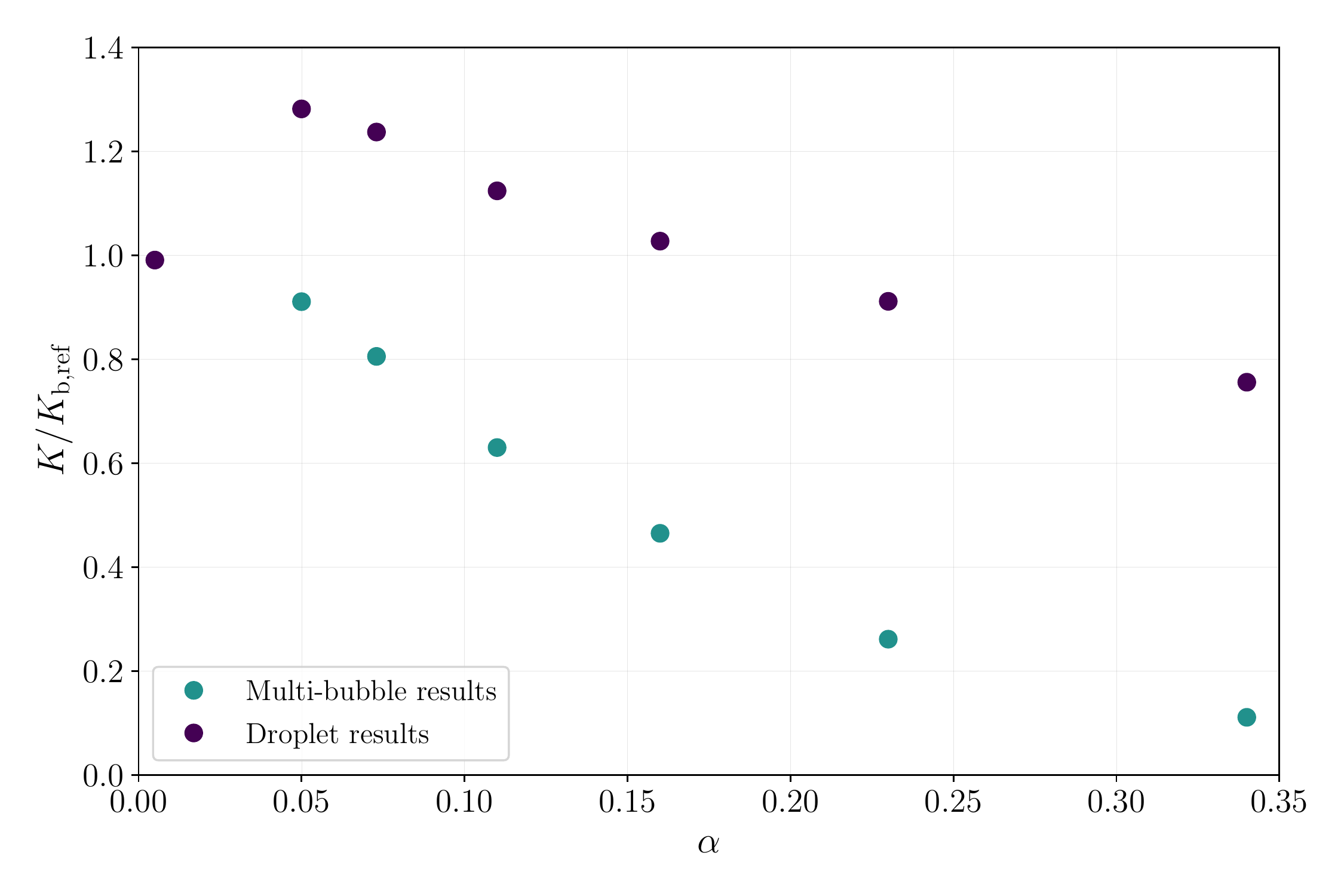}
  \caption{Comparison of the relative kinetic efficiency $K/\Kbref$ in
    the spherically symmetric 1D droplet simulations and the 3D
    multi-bubble simulations of Ref.~\cite{Cutting:2019zws}. The
    values of $K/\Kbref$ are taken from Table~\ref{tab:results}, see
    the caption for details.}\label{fig:K_scatter}
\end{figure}

Measuring the values for $\Kd/\Kbref$ at evaporation, we compare these
to equivalent values from 3D multi-bubble simulations. For the
multi-bubble simulations, we plot the maximum value of $K$ (see
Eq.~\ref{eq:KDefn}) from the simulation, and normalise to $\Kbref$
with $\Rbref=\Rstar/2$ and $\Rstar$ the mean bubble separation in the
simulation. The resulting plot is shown in
Fig.~\ref{fig:K_scatter}. We note that $K/\Kbref$ is consistently
larger for the spherical droplet simulations than in the 3D
multi-bubble simulations. However, there is a consistent downward trend
in both sets of simulations as $\Str$ is increased, with the exception
of $\Str = 0.005$ for the spherical droplets. Looking at the bottom
panel of Fig.~\ref{fig:rbtdot_K_wrt_time}, we note that the initial
value of $\Kd$ is significantly larger than $\Kbref$, with the value
for $\Str = 0.005$ noticeably smaller. This indicates that we can't
fairly compare our current droplet results to the multi-bubble results
because much more kinetic energy is produced from the initial
condition in the droplet case. On the other hand, as explained above,
if we did use the similarity solution as an initial condition, the
final $\Kd$ from a droplet would be close to zero. An accurate
estimation of the kinetic efficiency for droplets formed in a
multi-bubble simulation using spherical droplet simulations is
therefore unlikely, regardless of the initial conditions used.

\begin{table*}[t]
\begin{tabularx}{2\columnwidth} 
{X@{\hspace{0.2cm}} XX XX@{\hspace{-1.5cm}}}
\toprule \toprule \\
[-0.5em]
\multirow{3}{*}{$ \Str$} & 
\multicolumn{2}{l}
{$\qquad\qquad\quad\;\; \left|\vwass\right|$}  &
\multicolumn{2}{l}
{$\qquad\qquad\quad K/\Kbref$}
 \\ [0.5em] \cmidrule[\lightrulewidth](l{-0.2em}r{5.4em}){2-3} 
 \cmidrule[\lightrulewidth](lr{0.5em}){4-5} \\ [-0.5em]
& Droplet & Multi-bubble & Droplet & Multi-bubble
\\
[0.5em]
\midrule \\
0.0050& 0.23 & --- & 0.99 & --- 
\\
0.050& 0.20 & 0.22 & 1.3 & 0.91 
\\
0.073& 0.19 & 0.20 & 1.2 & 0.81 
\\
0.11 & 0.16 & 0.18 & 1.1 & 0.63 
\\
0.16 & 0.14 & 0.15 & 1.0 & 0.46 
\\
0.23 & 0.099 & 0.11 & 0.91 & 0.26 
\\
0.34 & 0.045 & 0.051 & 0.76 & 0.11 
\\
0.41 & 0.020 & --- & --- & --- 
\\[0.5 em] \bottomrule \bottomrule
\end{tabularx}
\caption{ Comparison of results from the spherically symmetric 1D
  droplet simulations and the 3D multi-bubble simulations of
  Ref.~\cite{Cutting:2019zws}.  We show the late-time wall speed
  $\left|\vwass\right|$ and the relative kinetic efficiency $K/\Kbref$
  for each transition strength.  The value of $\left|\vwass\right|$ is
  found from a fit to $\vws$ for $0.05 \RdI<\Rd<0.25 \RdI$ in the
  spherical droplet simulations, and for when the broken phase volume
  is between $10\%$ and $2\%$ in the multi-bubble simulations. See
  Appendix~\ref{app:wallspeed} for a discussion on wall speed
  estimators. The values of $K/\Kbref$ refer to the value of $\Kd$
  (see Eq.~\ref{eq:KdDefn}) at droplet evaporation in the spherical
  simulations, see black crosses in Fig.~\ref{fig:rbtdot_K_wrt_time},
  and to the peak value of $K$ (see Eq.~\ref{eq:KDefn}) in the
  multi-bubble simulations. $\Kbref$ (see Eq.~\ref{eq:KbDefn})
  corresponds to the kinetic energy of an isolated bubble with radius
  $\Rbref=\RdI$ for the spherical droplets and with $\Rbref=\Rstar/2$
  for the multi-bubble simulations.}\label{tab:results}
\end{table*}

\subsection{Self-similar droplets}\label{sec:self-similar}

In this section we explore whether any of the droplets in our
simulations reach a similarity solution before evaporating. We
qualitatively discuss the process by which a similarity solution forms
and its implications for multi-bubble simulations.

\begin{figure}[htbp]
  \centering
  \subfigure[]{\includegraphics[width=0.48\textwidth]{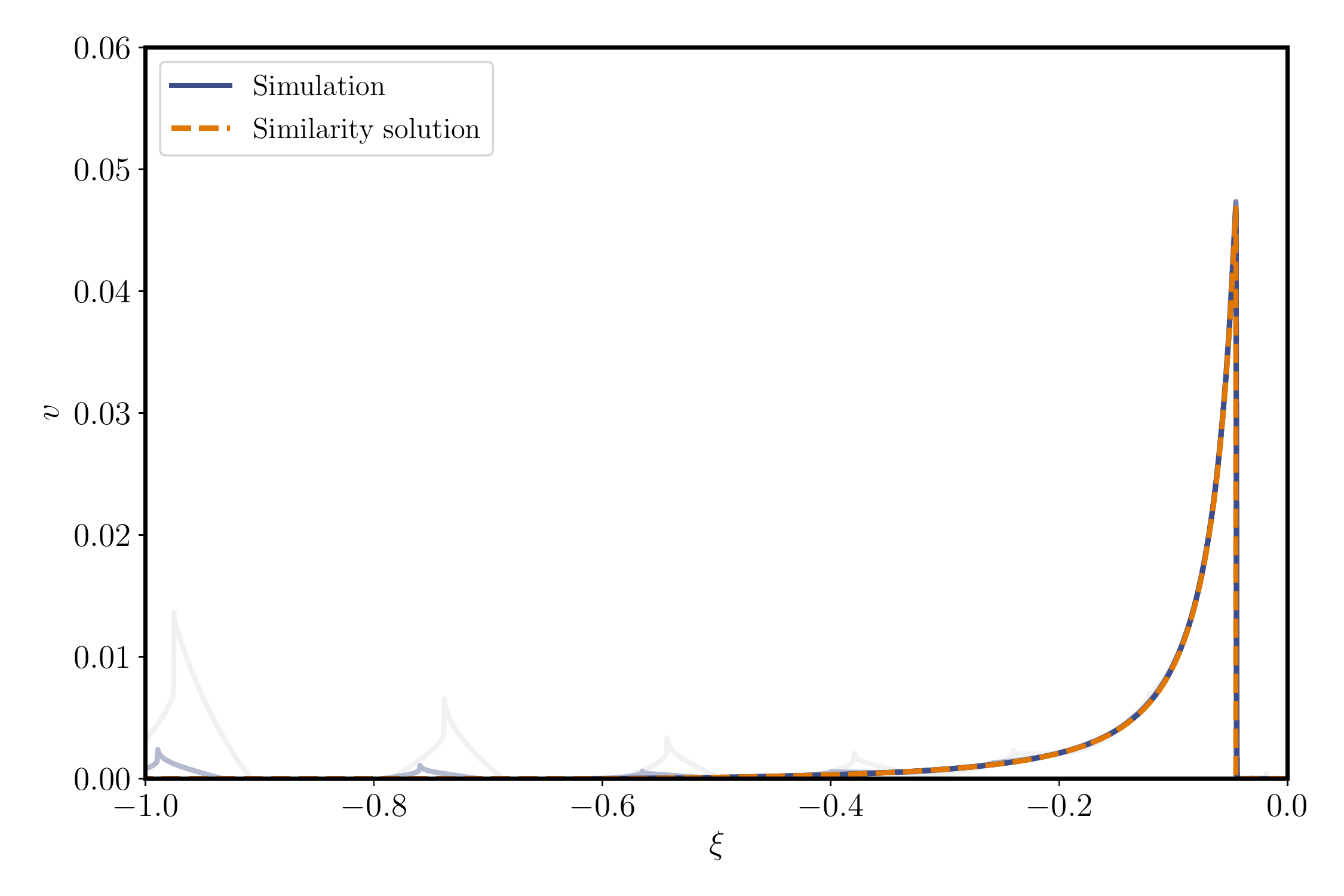}}
  \subfigure[]{\includegraphics[width=0.48\textwidth]{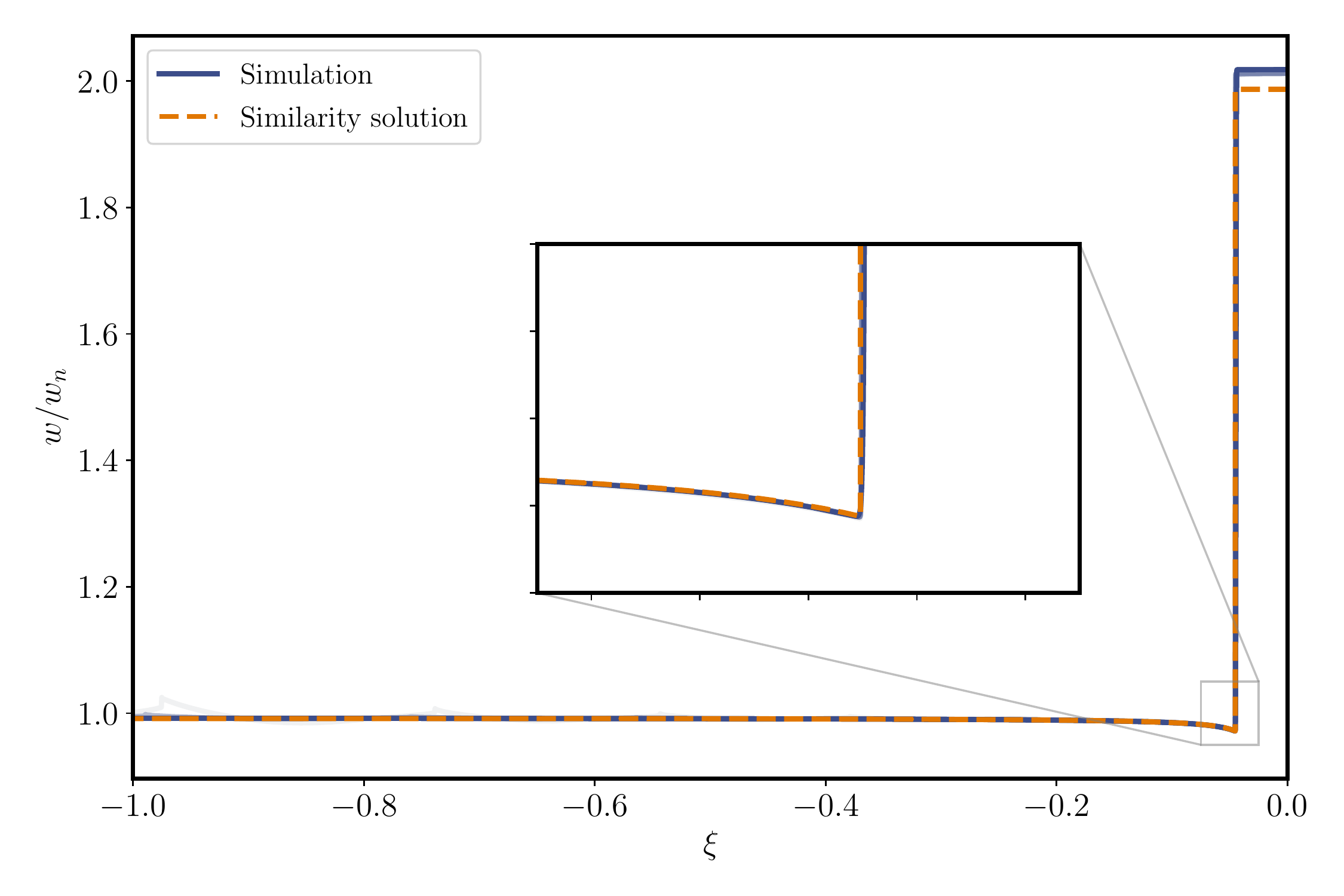}}
  \caption{Comparison between the late-time fluid profile in one of
    our spherical simulations and a self-similar profile. The upper
    plot~(a) shows the fluid velocity, while the lower plot~(b) gives
    the enthalpy $w$ normalised to the symmetric phase enthalpy at
    nucleation temperature $w_n \equiv w(0, \TN)$. The simulation
    shown here in blue has $\Str=0.34$. Six equally spaced snapshots
    of the simulation between $t\Tc=192\,000$ and $t\Tc=320\,000$ are
    shown in blue, with later snapshots shown using darker shades. The
    self-similarity variable $\xi=r/\ts$, where in this case $\ts =
    \Rd(t)/\vwass$ with $\vwass=-0.0452$. The dashed line corresponds
    to the similarity solution obtained by matching $v$ and $w$
    against the last timestep shown.\label{fig:SimilarityResults}}
\end{figure}

To begin with, we consider the droplet simulation with $\Str = 0.34$.
This appeared to display signs of approaching a similarity solution in
the rightmost column of Fig.~\ref{fig:lineoutalph0.34}.  We first need
to reconstruct $\xi =r/\ts$, where $\ts$ is defined such that
$\ts\Tc=0$ corresponds to the evaporation of the droplet (see the
detailed discussion in Section~\ref{sec:similarity}).  To compare the
result of our simulation with the similarity profile, we take $\ts =
\Rd(t)/\vw(t)$, where $\vw$ is negative for a contracting droplet. The
velocity and enthalpy profiles in the simulation at late times are
plotted as a function of $\xi$ in Fig.~\ref{fig:SimilarityResults}. We
also plot the similarity curves found by integrating
Eqs.~(\ref{eq:sim-xi}-\ref{eq:sim-w}) starting from $\xi = \vw$ and
with the initial velocity (enthalpy) taken from the maximum (minimum)
of the corresponding simulation profile. The predicted enthalpy for
the similarity solution inside the droplet can be found from
Eq.~(\ref{eq:sim-match2}). From Fig.~\ref{fig:SimilarityResults} we
can see that there is good agreement at late times for the droplet
simulation with $\Str = 0.34$ and a similarity solution.

A self-similar solution has the fluid at rest inside the
droplet\footnote{Fluid moving inside the droplet would impose a length
scale, breaking the self-similarity; it would also have to fit into an
ever smaller space as the droplet collapses and
$\ts\Tc\to0^{-}$.}. Therefore, for a droplet to become self-similar,
the fluid perturbations arising from the inward-propagating shell
created by the initial conditions must first exit the droplet.  Where
we observe this to occur, the process takes place over multiple
collisions of the fluid perturbations with the phase boundary and the
origin.  In each collision with the phase boundary, part of the fluid
perturbation is transmitted to the exterior of the droplet.
Sufficient time is therefore required for the fluid perturbations to
make multiple repeat journeys between the phase boundary and the
origin before the droplet evaporates.

As can be seen from Fig.~\ref{fig:rbtdot_K_wrt_time}, stronger phase
transitions cause the phase boundary to propagate at a reduced speed,
effectively delaying the evaporation of the droplet and allowing the
interior of the droplet to settle to a constant temperature, with the
fluid at rest. It is not clear if self-similar solutions would be
obtained for the weaker transitions we consider if $\RdI$ were to be
increased; we leave this for future study.

\section{Conclusions}\label{sec:Conclusions}

In this work, we studied the collapse of spherically symmetric
droplets in the coupled field--fluid model with a bag-like equation of
state. We found the value of $\eta$ that corresponded to an asymptotic
expanding-bubble wall speed of $\xiw=0.24$ for a range of transition
strengths. This value of $\eta$ was then used to perform a collapsing
droplet simulation.

Our initial conditions consisted of an at-rest tanh-like profile for
the scalar field, with the fluid at rest and at the nucleation
temperature everywhere. Upon release, an inward-moving fluid shell
develops in front of the phase boundary. The fluid shell rebounds upon
collision with the origin or with the phase boundary, which can happen
multiple times.  During a collision with a fluid shell, the phase
boundary is decelerated and can even temporarily reverse
direction. This effect is more pronounced at higher transition
strengths.

The late-time wall velocity for these spherically symmetric droplets
approximately agrees with the late-time wall speed found in the
multi-bubble simulations, despite the droplets of
Ref.~\cite{Cutting:2019zws} not being spherical. This possibly
indicates that the wall velocity is determined by the local conditions
across the phase boundary, rather than the large-scale geometry of the
droplet itself. On the other hand, the kinetic energy efficiency in
the droplets does not match that extracted from multi-bubble
simulations, although the same downward trend in the kinetic
efficiency as $\Str$ increases is observed in both cases. The
discrepancy between the simulations is presumably caused by a burst of
kinetic energy associated with the generation of the initial fluid
shell in the spherical droplet case.

For the strongest transitions, the droplet approaches a
self-similarity solution at late times, whereas for our weak and
intermediate transitions this does not occur. The approach to a
self-similarity solution requires the fluid perturbations to first
exit the droplet.  In our simulations, this only occurs for strong
transitions in which the phase boundary slows significantly. This
allows for repeated collisions of the fluid perturbations with the
phase boundary and for the perturbations to eventually leave the
droplet.  Despite this, it appears that the late-time wall velocity
found in droplets is more generic than one might expect, as the wall
speed in the spherical simulations seem to approximately match the
multi-bubble simulations regardless of whether a self-similar solution
develops.

In the case of the strongest transition we considered, the magnitude
of the droplet wall velocity drops to less than $10\%$ of that of an
isolated expanding bubble with the same value of $\eta$. This could
modify baryogenesis predictions for strong transitions, as the baryon
asymmetry that is generated during a phase transition is strongly
dependent on the wall velocity (see
Refs.~\cite{Cline:2020jre,Azatov:2021irb,Baldes:2021vyz,Dorsch:2021ubz,Lewicki:2021pgr}
for recent results on this phenomenon).

While the kinetic energy efficiency disagrees between the spherical
droplet and multi-bubble simulations, in both cases we saw a decrease
in $K/\Kbref$ as $\Str$ increased. For a self-similarity solution,
$\Kd$ tends towards zero as the droplet shrinks. From the spherical
simulations, we saw that stronger transitions have longer to relax
towards a similarity solution due to the deceleration of the phase
boundary. It is possible that this is also occurring in the
multi-bubble simulations, and that we are seeing the same effect
obscured by the fluid perturbations induced by the different initial
conditions.

An interesting question is whether a self-similarity solution would
eventually develop for a droplet that is produced in a multi-bubble
collision. It seems unlikely that this would occur, as not only are
the initial droplets non-spherical, but also in a realistic phase
transition there are long-lasting fluid perturbations that propagate
in all directions. These fluid perturbations would prevent the
interior of the droplet from completely relaxing, which in turn
prevents the similarity solution from fully developing.

Our results indicate that important phase transition quantities such
as the late-time wall speed can be computed from spherical
simulations, reducing the need for expensive 3D multi-bubble
simulations. Other quantities, like the kinetic energy production, are
harder to match onto multi-bubble simulations. One quantity we have
not discussed in this work is the fraction of the universe in which
droplets are formed. If this fraction could also be estimated using
spherical simulations, it could be used in conjunction with the
late-time wall speed to provide an estimate for the enhancement factor
for the baryon asymmetry. If the kinetic energy efficiency factor for
droplets could also be estimated, then a suppression factor for
gravitational wave production could be determined using spherical
simulations.

\begin{acknowledgments}
  We acknowledge useful discussions with Mark Hindmarsh, Asier
  Lopez-Eiguren and Kari Rummukainen, and thank Oliver Gould for
  helpful comments on an earlier version of this
  manuscript. D.C. (ORCID ID 0000-0002-7395-7802) was supported by
  Academy of Finland grant nos. 328958 and 345070. E.V. (ORCID ID
  0000-0002-5240-5865) was supported by the Research Funds of the
  University of Helsinki and Academy of Finland grant
  no. 328958. D.J.W. (ORCID ID 0000-0001-6986-0517) was supported by
  Academy of Finland grant nos. 324882 and 328958. The authors would
  also like to thank Finnish Grid and Cloud Infrastructure at the
  University of Helsinki (urn:nbn:fi:research-infras-2016072533) and
  CSC – IT Center for Science, Finland, for computational
  resources. We acknowledge PRACE for awarding us access to HAWK at
  GCS@HLRS, Germany.
\end{acknowledgments}

\bibliography{spherical-droplet-biblio}

\newpage
\appendix
\section{Wall speed estimators}\label{app:wallspeed}

The wall speed can be a challenging quantity to estimate. For isolated
droplets and bubbles in our the spherical simulations, the wall speed
can be calculated directly from the position of the phase
boundary. However, the situation becomes more complicated for 3D
multi-bubble simulations.

In the spherical simulations, we find the wall velocity by tracking
the midpoint of the phase boundary. Throughout the simulation we
regularly find the position $R$ for which $\phi(R)= \phib/2$. The wall
velocity $\vw$ is then found via the time derivative of $R$ using a
first-order forward difference
\begin{equation}
\vw(t) \approx \frac{R(t+N\Delta t) - R(t)}{N\Delta t}\text,
\end{equation}
where $N$ refers to the number of timesteps between successive outputs
of the position $R$.

To find an approximate measure of the wall speed $\vws$ in the 3D
multi-bubble simulations of Ref.~\cite{Cutting:2019zws}, we use two
different methods. The first approach is to measure the rate of change
of the volume in the broken phase and the area of the phase
boundary. On the lattice we approximate the volume in the broken phase
to be
\begin{equation}
  \mathcal{V}_\mathrm{broken} = (\Delta x)^3 \sum_{\mathbf{n}}
  \begin{cases}1\text{ if }\phi_\mathbf{n} \geq \phi_b/2\text, \\ 
    0\text{ otherwise.} \\ 
  \end{cases}
\end{equation}
where $\mathbf{n}$ denotes the lattice coordinate vector and the
summation is over the whole lattice.  To find the surface area of the
phase boundary we use
\begin{align}
  A = \frac{2}{3}(\Delta x)^2 \sum^N_\mathbf{n}&\left[ f(\phi_\mathbf{n},\phi_{\mathbf{n} + \hat\imath})\right.
                                                 +f(\phi_\mathbf{n},\phi_{\mathbf{n} + \hat\jmath}) \nonumber \\
                                               &+\left.f(\phi_\mathbf{n},\phi_{\mathbf{n} + \hat k}) \right]\text,
\end{align}
where the function
\begin{equation}
  f(\phi,\phi') =
  \begin{cases}1\text{ if }(\phi - \phib/2)(\phi' - \phib/2) < 0\text, \\ 
    0\text{ otherwise.}
  \end{cases}
\end{equation}
checks whether the field crosses $\phib$ between two sites. The factor
of $2/3$ compensates for the over counting of a the surface area of a
sphere represented on a cubic grid in the asymptotic limit of
infinitesimally small grid spacing \cite{OEIS}.  The average wall
speed $\vws$ can then be approximated by
\begin{equation}
  \label{eq:vwvolume}
  \vws \approx \frac{1}{A}\frac{\mathrm{d}
    \mathcal{V}_\mathrm{broken}}{\mathrm{d}t}.
\end{equation}

We can alternatively compute the wall speed from the kinetic and
gradient energies in the scalar field. Assuming that at the phase
boundary the scalar field obeys a transport equation
\begin{equation}
\partial_t \phi - \mathbf{\vw} \cdot \nabla \phi = 0\text,
\end{equation}
the average wall speed can then be estimated using
\begin{equation} \label{eq:vwenergies}
\vws \approx \sqrt{\frac{E_K}{E_D}}\text,
\end{equation}
where $E_K$ and $E_D$ are the average kinetic and gradient energy
densities in the simulation, respectively. We have made use of the
fact that the scalar field varies only at the phase boundary.

\begin{figure}[htbp]
  \centering
\includegraphics[width=0.48\textwidth]{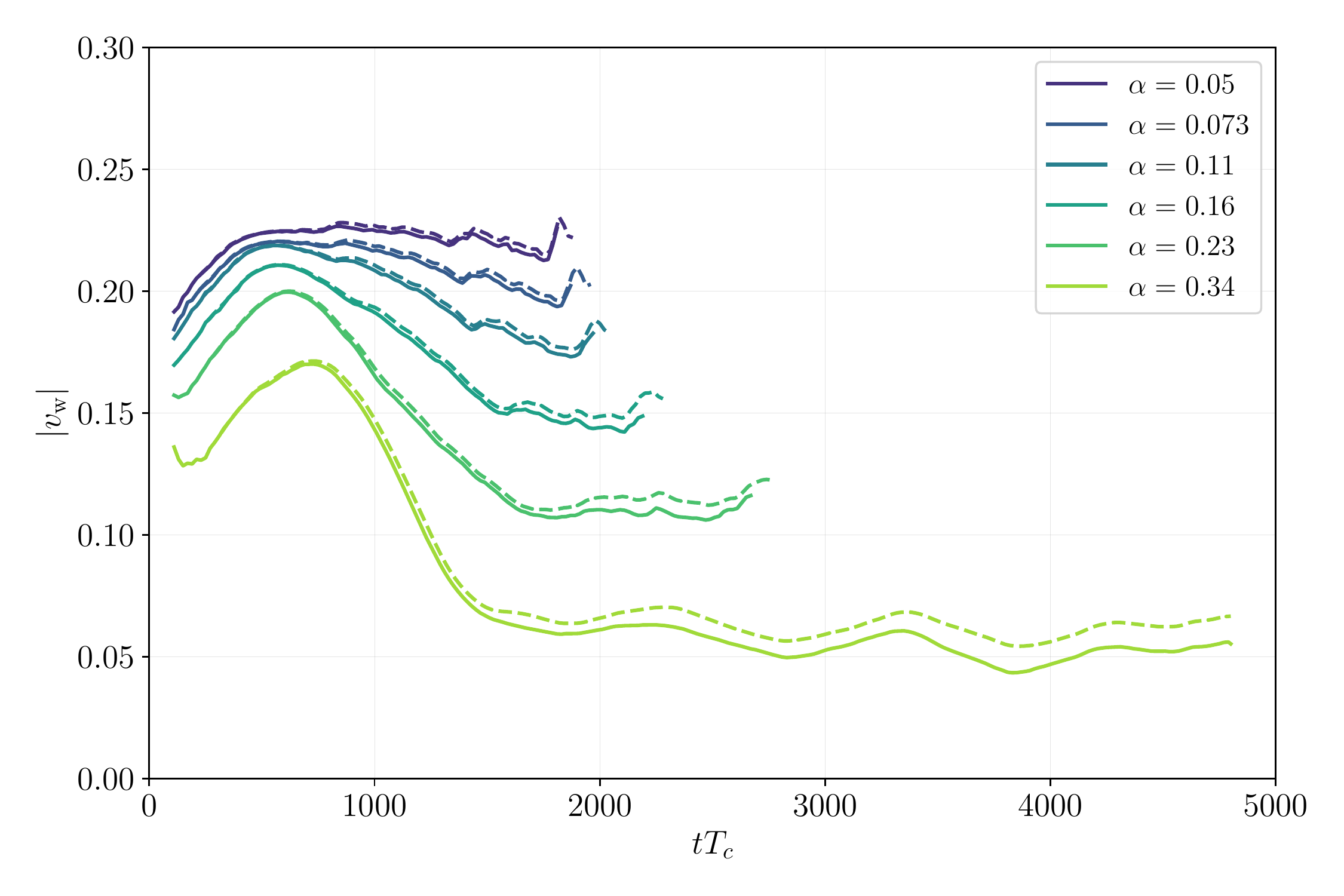}
\caption{Comparison of the two different wall speed estimators
  discussed in this appendix (c.f. Fig.~\ref{fig:3DWallSpeed}). The
  solid line corresponds to the wall speed as estimated using the
  broken phase volume and phase boundary surface area (c.f.
  Eq.~(\ref{eq:vwvolume})), whereas the dashed line gives the speed as
  estimated using the kinetic and gradient energies (c.f.
  Eq.~(\ref{eq:vwenergies})).\label{fig:3DWallSpeedMethodComp}}
\end{figure}

We compare these two methods in
Fig.~\ref{fig:3DWallSpeedMethodComp}. Both velocity estimators agree
for the majority of the duration of all the simulations. The largest
disagreement is at late times for the simulation with $\Str = 0.34$,
but even here the estimator from the kinetic and gradient energies
differs by at most $20\%$ from the estimator using the broken phase
volume and boundary surface area. We use the broken phase volume and
boundary surface area estimator in the main body of this paper.

\section{Effect of varying the fractional change in the degrees of freedom}
\label{app:fractionalchange}

\begin{figure*}[htbp]
  \centering
  \subfigure[\,$\alpha=0.05$]{\includegraphics[width=0.98\textwidth]{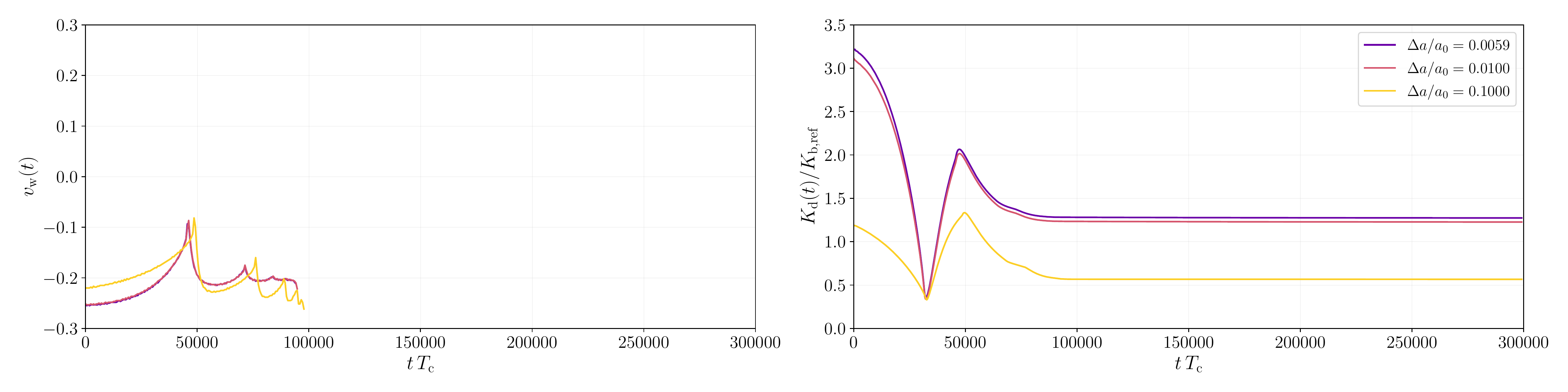}} \\
  \subfigure[\,$\alpha=0.11$]{\includegraphics[width=0.98\textwidth]{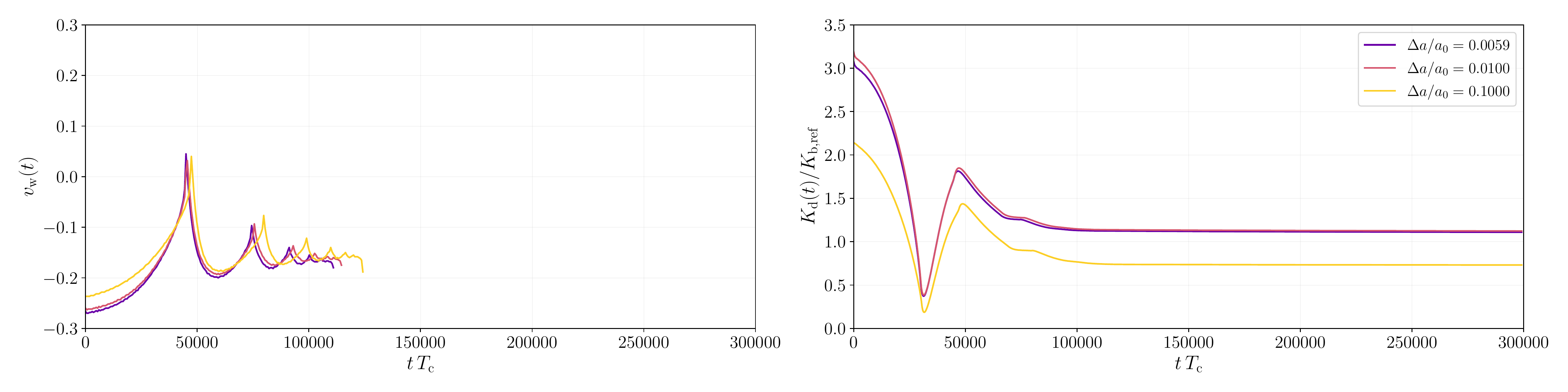}} \\ 
  \subfigure[\,$\alpha=0.16$]{\includegraphics[width=0.98\textwidth]{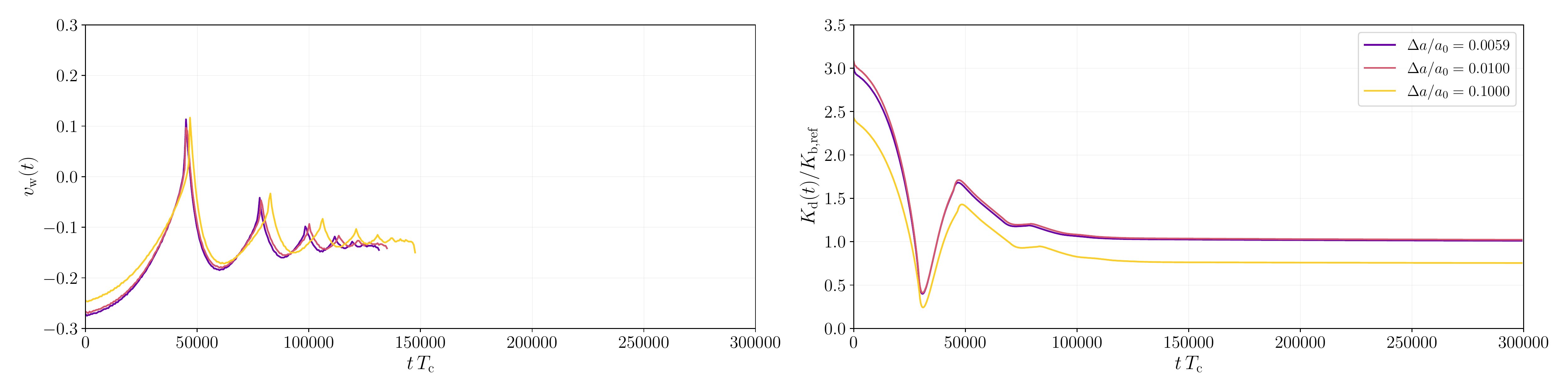}} \\
  \subfigure[\,$\alpha=0.34$]{\includegraphics[width=0.98\textwidth]{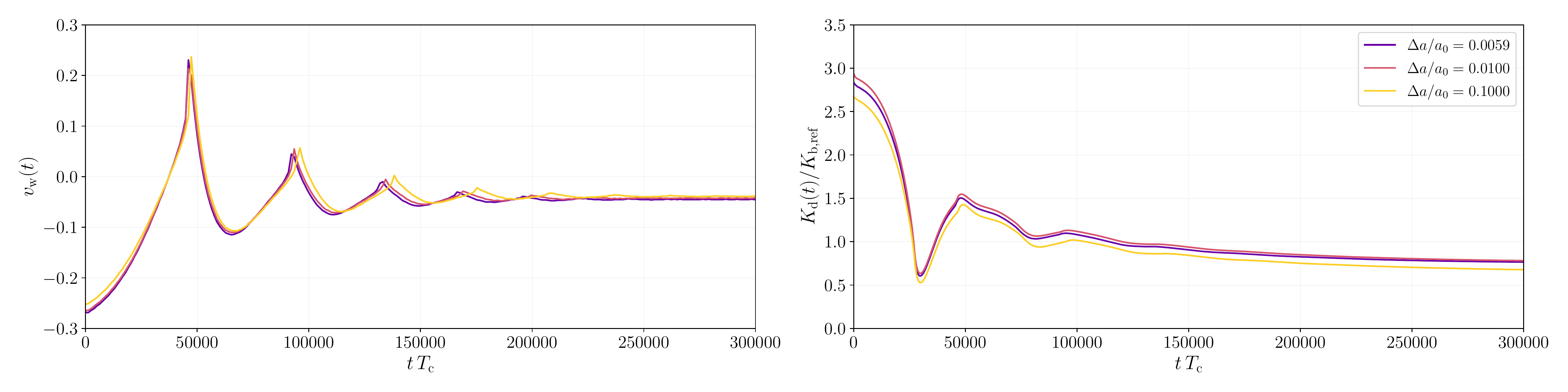}}
  \caption{Plots showing the effect of varying the fractional change
    in degrees of freedom, $\Delta a/a_0$ on the wall velocity $\vw$
    and normalised kinetic energy fraction $\Kd/\Kbref$. Four choices
    of $\alpha$ are shown, with the same three choices of $\Delta
    a/a_0$ shown for each. To allow comparison, the axes are the same
    for all choices of $\alpha$; the $\alpha=0.34$ case is therefore
    cut off at $t = 300000/\Tc$, before the droplets evaporate. As
    $\alpha$ is increased, the effect of varying $\Delta a/a_0$
    becomes milder. \label{fig:vary_dofchange}}
\end{figure*}

In the main body of this paper, we focus on a particular choice of the
change in the number of light degrees of freedom $\Delta a/a_0 \approx
0.0059$. One might expect that stronger transitions are easier to
achieve with larger changes in the number of light degrees of freedom;
for example, they will in general require less supercooling to achieve
a comparable transition strength. It is therefore worth exploring the
extent to which the results we present in the main body of this paper
depend on our choice of $\Delta a/a_0$.
 
In Fig.~\ref{fig:vary_dofchange} we plot the wall velocity and
$\Kd/\Kbref$ as a function of time for several phase transition
strengths $\Str$, for a variety of choices of the fractional change in
the number of degrees of freedom across the transition, including
$\Delta a/a \approx 0.0059$ used in the main body of the paper.

In varying $\Delta a/a$, we keep $M^2$ and $\lw$ constant. We allow
$\mu$ and $\lambda$ to change and then adjust $\TN$ and $\eta$ to
achieve the desired $\alpha$ and $\xiw$.

For large $\Str$ we see good agreement between the different choices
of $\Delta a/a_0$ for the observables of interest. Therefore, for
strong transitions, the droplet behaviour we see in the main body of
the paper is expected to occur, independent of the choice of $\Delta
a/a_0$.

For smaller $\Str$ there is some disagreement, with larger changes in
the number of degrees of freedom beginning with a shift in the amount
of kinetic energy generated at early times. We note that, with our
equation of state, the pressure difference across the wall at $\TN$ is
  \begin{equation}
    \Delta p \equiv p(\TN,0) - p(\TN,\phib) = -\frac{\Delta a}{a_0} \left( 1- \frac{\Delta a}{a_0} \frac{1}{3\alpha}\right) a_0 T_c^4.
  \end{equation}
The magnitude of the pressure difference is approximately linear in
$\Delta a/a_0$ for sufficiently large $\alpha$, but it decreases for
$\Delta a/a_0 > 3 \alpha/2$.

The initial pressure difference determines the initial wall velocity
-- even though we have adjusted $\eta$ so that an expanding bubble
would reach the same asymptotic wall velocity $\xiw$. A slower initial
wall velocity leads to less kinetic energy transfer, which has
long-lasting consequences for $\Kd/\Kbref$. We believe that this
explains the results for $\Delta a/a_0 \gtrsim \alpha$ seen in
Fig.~\ref{fig:vary_dofchange}.
  
\section{Convergence with initial droplet radius}
\label{app:convergence}

\begin{figure}[htbp]
  \centering
  \includegraphics[width=0.48\textwidth]{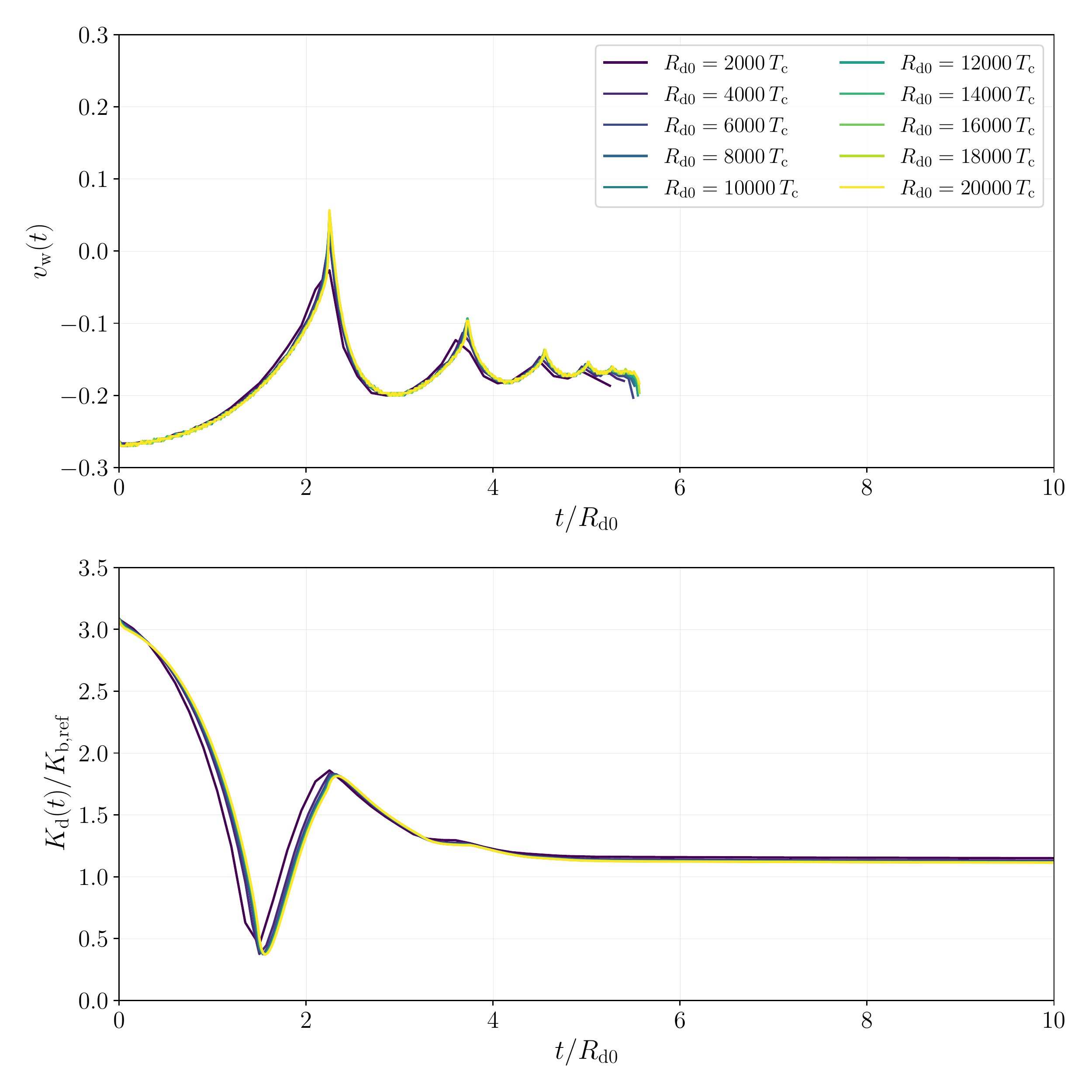}
  \caption{Plot showing the effect of the initial droplet radius on
    the results of interest in the main body of the paper, with
    $\alpha=0.11$. The wall velocity $\vw$ and droplet radius $\Kd$
    normalised to the initial bubble radius $\Kbref$ are both shown,
    as a function of time normalised to initial droplet radius
    $\RdI$. The curves generally collapse onto a single line, showing
    minimal dependence on the initial droplet radius. As in
    Fig.~\ref{fig:rbtdot_K_wrt_time}, the lines in the upper plot end
    when the phase boundary reaches the origin. The measurement
    interval was kept constant at 1500 timesteps, leading to some
    aliasing in the results.\label{fig:vary_radius}}
\end{figure}

In Fig.~\ref{fig:vary_radius} we show how the wall velocity and
$\Kd/\Kbref$ vary with time for a variety of different initial droplet
radii $\RdI$. We normalise the time by the initial radius, and find
that for initial radii over one order of magnitude the results
collapse onto a single line. Note that the reference radius taken for
$\Kbref$ is, in line with the choice made in
Section~\ref{sec:Results}, the initial droplet radius $\RdI$ and hence
is different for each case shown.

The only meaningful combination of length scales in the initial
conditions is the bubble wall width relative to the initial radius. In
the early universe, the wall width will likely be many orders of
magnitude smaller than the radius of a hot droplet, given the
difference in scale between the bubble wall width $\lw$ and the
typical distance between bubbles $\Rstar$. The collapse of these
curves onto a single line gives us confidence that our results can be
extrapolated to the physical case, where the separation is potentially
much larger.

\end{document}